\documentclass{llncs}

\frontmatter         
\usepackage{array,booktabs,amsmath,amssymb,stmaryrd}
\usepackage[english]{babel}
\usepackage[breaklinks]{hyperref}
\usepackage{breakurl}
\usepackage{xcolor,xspace}
\usepackage{enumerate,wrapfig}

\usepackage[obeyDraft]{todonotes} 
\newcommand{\todoeasy}[1]{\todo[color=green]{#1}}
\newcommand{\todocomment}[1]{\todo[color=gray]{#1}}
\newcommand{\todowarn}[1]{\todo[color=yellow]{#1}}
\newcommand{\todohard}[1]{\todo[color=pink]{#1}}
\newcommand{\comment}[1]{}

\newif\ifunused

\usepackage{caption} 
\usepackage[final]{listings}
\lstdefinestyle{customc}{
  breaklines=true,
  frame=l,
  framesep=1pt,
  xleftmargin=\parindent,
  language=C,
  numbers=left,
  numbersep=2pt,
  basicstyle=\scriptsize\ttfamily,
  keywordstyle=\color{blue}, 
  commentstyle=\color{purple!40!black}
}

\usepackage{tikz}
\usetikzlibrary{positioning}
\usetikzlibrary{shapes.symbols}

\newcommand{\code}[1]{\mbox{\lstinline[language=C,basicstyle=\ttfamily]{#1}}}

\newcommand{\eg}{e.g.}
\newcommand{\ie}{i.e.}
\newcommand{\resp}{resp.}

\newcommand{\mypar}[1]{\smallskip\noindent\emph{#1}}

\newcommand{\csyn}[1]{\mathtt{#1}}
\newcommand{\carith}{\mathtt{num}}
\newcommand{\cinttyps}{\mathtt{num}}

\newcommand{\cptr}[1]{#1~\mathtt{ptr}}
\newcommand{\cptrtyps}{\mathsf{Ptr}}
\newcommand{\carr}[2]{#1\mathtt{\lbrack}#2\mathtt{\rbrack}}

\newcommand{\ctyps}{\mathtt{Ctyp}}  
\newcommand{\cty}{\mathtt{t}}
\newcommand{\cstyps}{\mathtt{Styp}}

\newcommand{\csty}{\mathtt{u}}
\newcommand{\cops}{\mathtt{O}}      
\newcommand{\cop}{\mathtt{op}}
\newcommand{\crecs}{\mathtt{Rtyp}}  
\newcommand{\crec}{\mathtt{rt}}
\newcommand{\cflds}{\mathtt{Fld}}  
\newcommand{\cfld}{\mathtt{f}}
\newcommand{\cvars}{\mathtt{Cvar}}  
\newcommand{\cvar}{\mathtt{v}}
\newcommand{\cexps}{\mathtt{Expr}}  
\newcommand{\cexp}{\mathtt{e}}
\newcommand{\ciexps}{\mathtt{Iexpr}}  
\newcommand{\ciexp}{\mathtt{ie}}
\newcommand{\caexps}{\mathtt{Aexpr}}  
\newcommand{\caexp}{\mathtt{a}}
\newcommand{\clvals}{\mathtt{Lval}} 
\newcommand{\clval}{\mathtt{lv}}
\newcommand{\cstmts}{\mathtt{Stmt}} 
\newcommand{\cstmt}{\mathtt{s}}
\newcommand{\cnull}{\mathtt{null}}

\newcommand{\csizeof}{
\mathtt{sizeof}}
\newcommand{\cassert}{
\mathtt{assert}}

\newcommand{\NN}{\mathbb{N}}  
\newcommand{\ZZ}{\mathbb{Z}}  
\newcommand{\pset}[1]{2^{#1}} 
\newcommand{\ft}{\rightarrow}  
\newcommand{\fst}{\textit{fst}} 
\newcommand{\snd}{\textit{snd}} 

\newcommand{\sem}[1]{{[\![\,#1\,]\!]}} 
\newcommand{\mathsem}[1]{\llbracket#1\rrbracket}

\newcommand{\cfoff}[1]{\mathtt{offset} (#1)}     
\newcommand{\ctsz}[1]{\mathtt{sizeof}(#1)}         
\newcommand{\ctyof}[1]{\mathtt{cty} (#1)}        
\newcommand{\ctyto}[2]{\mathtt{cast} (#1,#2)} 
\newcommand{\typeof}{\mathtt{cty}}

\newcommand{\ammem}{\textit{Mem}}     
\newcommand{\amv}{\textit{m}}           
\newcommand{\amblk}{\textit{Block}}   
\newcommand{\amloc}{\textit{Loc}}     
\newcommand{\amval}{\textit{Val}}     
\newcommand{\amvint}{\textit{Vint}} 
\newcommand{\amvptr}{\textit{Vptr}} 

\newcommand{\ambase}{\textit{base}}   
\newcommand{\amshft}{\textit{shift}}  
\newcommand{\amload}{\textit{load}}   
\newcommand{\amstore}{\textit{store}} 

\newcommand{\vint}[1]{\mathtt{Vint} \left(#1\right)} 
\newcommand{\vptr}[1]{\mathtt{Vptr} \left(#1\right)} 

\newcommand{\vcgen}[1]{\textsf{#1}}
\newcommand{\mmmem}{\textsf{Mem}}     
\newcommand{\mmv}{\textsf{m}}           
\newcommand{\mmloc}{\textsf{Loc}}     
\newcommand{\mml}{\textsf{l}}         
\newcommand{\mmval}{\textsf{Val}}     
\newcommand{\mmvint}{\textsf{Vint}}   
\newcommand{\mmvptr}{\textsf{Vptr}}   

\newcommand{\mmbase}{\textsf{base}}   
\newcommand{\mmshft}{\textsf{shift}}  
\newcommand{\mmshift}{\textsf{shift}}  
\newcommand{\mmload}{\textsf{load}}   
\newcommand{\mmstore}{\textsf{store}} 

\newcommand{\compile}[1]{\langle #1 \rangle}

\newcommand{\logic}[1]{\underline{#1}}

\newcommand{\llogics}{\mathcal{T}} 
\newcommand{\ltheory}{\mathcal{T}} 
\newcommand{\lterms}{\mathcal{E}}  
\newcommand{\lvars}{\mathcal{X}}   

\newcommand{\lbools}{\mathbb{B}}    
\newcommand{\lintegers}{\mathbb{I}} 
\newcommand{\larray}{\ensuremath{\mathit{array}}}

\newcommand{\mathabs}[1]{#1^{\#}}
\newcommand{\abstr}{^\#}

\newcommand{\absLoc}{\mathabs{Loc}}
\newcommand{\absloc}{\mathabs{\ell}}

\newcommand{\Cells}{\mathcal{C}}
\newcommand{\CellRange}{r}

\newcommand{\mmvdfont}{\mathsf}

\newcommand{\mmvdloc}{\mathcal{L}}
\newcommand{\mmvdst}{\mathcal{S}}
\newcommand{\mmvdblk}{\mathcal{B}}

\newcommand{\mmvdload}{\mmvdfont{load}}
\newcommand{\mmvdsli}{\mmvdfont{slice}}
\newcommand{\mmvddom}{\mmvdfont{domain}}

\newcommand{\Value}{\textsf{Eva}\xspace}
\newcommand{\WP}{\textsf{WP}\xspace}
\newcommand{\Qed}{\textsf{Qed}\xspace}

\newcommand{\framac}{\textsf{Frama-C}\xspace}

\newcommand{\clight}{\textsf{Clight}\xspace}

\newcommand{\er}{\textbf}
\newcommand{\erival}{\er{ilvl}}

\newcommand{\ertyped}{\er{Typed}}
\newcommand{\erbase}{\er{B}}
\newcommand{\erbasetop}{\erbase$_\top$}
\newcommand{\erpa}{\er{P}}
\newcommand{\ercell}{\er{C}}

\newcommand{\ersigs}{\er{N}}

\newcommand{\ervars}{\er{vars}}
\newcommand{\erarrays}{\er{arrs}}
\newcommand{\erstruct}{\er{strct}}
\newcommand{\erreg}{\er{grp}}
\newcommand{\errdm}{\er{rdn}}

\newcommand{\erpamax}{\erpa_{\text{max}}}
\newcommand{\erpafew}{\erpa_{\text{few}}}

\begin{document}

\title{Exploiting Pointer Analysis in Memory Models for Deductive
  Verification\todoeasy{Ideas? see notes}{} 
}

\author{%
    Quentin Bouillaguet\inst{1,2}
\and\ 
    Fran{\c{c}}ois Bobot\inst{1}
\and\\
    Mihaela Sighireanu\inst{2} 
\and\ 
    Boris Yakobowski\inst{1,3}
}

\institute{%
    CEA, LIST, Software Reliability Laboratory, France \\
\and
    IRIF, University Paris Diderot and CNRS, France \\
\and
    AdaCore, Paris, France \\
}

\authorrunning{Q. Bouillaguet et al.}

\sloppy
\maketitle

\begin{abstract}
Cooperation between verification methods is crucial to tackle the challenging problem of software verification. 
The paper focuses on the verification of C programs using pointers and 
it formalizes a cooperation between static analyzers doing pointer analysis and a deductive verification tool based on first order logic.
We propose a framework based on memory models that captures the partitioning of memory inferred by pointer analyses, and complies with the memory models used to generate verification conditions.
The framework guided us to propose a pointer analysis that accommodates to various low-level operations on pointers 
    while providing precise information about memory partitioning to the deductive verification.
We implemented this cooperation inside the \framac\ platform and 
we show its effectiveness in reducing the task of deductive verification
on a complex case study.\todowarn{Keep less 150 words!}{} 

\end{abstract}


\section{Introduction}

Software verification is a challenging problem for which different
solutions have been proposed.
Two of these solutions are deductive verification (DV) and static analysis (SA).

Deductive verification is interested in checking precise and expressive properties of the input code. It requires efforts from the user that has to specify the properties to be checked, plus other annotations -- e.g., loop invariants. Using these specifications, DV tools build verification conditions which are formulas in various logic theories and send them to specialized solvers.
For C programs with pointers, DV has been boosted by the usage of Separation Logic~\cite{OHearnRY01}, which leads to compact proofs due to the local reasoning allowed by the separating conjunction operator. 
However, for programs with low-level operations on pointers (\eg, pointer arithmetics and casting), this approach is actually limited by the theoretical results on the fragment of separation logic employed~\cite{BrotherstonK18} and on the availability of solvers.
Therefore, this class of programs is most commonly dealt using classic approaches based on memory models \emph{\`a la} Burstall-Bornat~\cite{ma/Burstall72,Bornat00}, which may be adapted to be sound in presence of low-level operations~\cite{RakamaricH09} and dynamic allocation~\cite{DBLP:conf/vmcai/0062BW17}.
The memory model is chosen in general by the DV engine which may employ some heuristics to guide the choice~\cite{hubert07hav}. Indeed, changing the memory model may result in an increase of the number of proofs discharged automatically~\cite{TuchKN07}.
However, annotations on non aliasing between pointers and memory partitioning complicates the task of users and of underlying solvers.

On the other hand, static analysis targets checking a fixed class of properties. This loss in the expressivity of properties is counterbalanced by a high degree of automation. 
For example, static pointer analysis for C programs usually computes over-approximations of the set of values (addresses) for each pointer expression at each control point. These abstractions do not speak about concrete memory addresses, but refer to symbolic memory regions provided by the memory allocated to program variables and in heap by dynamic allocation methods.

The information obtained by static analysis may help to infer partitioning of the memory in disjoint regions which can then be used by DV tools.
The success of this collaboration between SA and DV tool strongly depends on the coarseness of the abstraction used by SA to keep track of the locations scanned by a pointer inside each memory region. 
For example, consider \code{p} a pointer to integer and a variable \code{s} of type record with five integer fields, \code{struct \{int m,n,o,p,q;\}}, 
such that \code{p} scans locations of all fields of \code{s} except \code{o} 
(i.e., \code{&s.m}, \code{&s.n}, \code{&s.p} and \code{&s.q}). 
Pointer analyses (e.g.,~\cite{lctrts/Mine06}) over-approximate the location of \code{p} to any location in the memory region of \code{s} which is multiple of an integer, thus including the spurious \code{o} field.
Therefore, it is important to be able to try several SA algorithms to gather precise information about the memory partitioning.

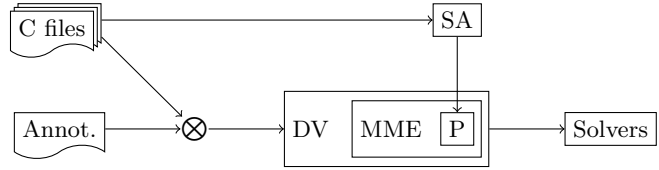
\begin{figure}[t]
\begin{center}
\newcommand{\cfile}[3]{\node[tape, draw, fill=white, tape bend
  top=none] (#3) at (#1, #2) {C files};}
\begin{tikzpicture}
  \cfile{0.1}{0.1}{cfile}
  \cfile{0.05}{0.05}{cfile2}
  \cfile{0}{0}{cfile3}
  \node[tape, draw, fill=white, tape bend top=none, below=of cfile] (annot) {Annot.};
  \node[inner xsep = 0,inner ysep = 0,right=of annot] (bigotimes) {$\bigotimes$};
  \node[minimum height=1cm,right=of bigotimes,draw] (dv) {\parbox{25mm}{DV}}; 
  \node[minimum height=7.5mm,at=(dv.east),draw,left=1mm] (m) {\parbox{15mm}{MME}}; 
  \node[at=(m.east),draw,left=1mm,draw] (p) {P};
  \node[at=(cfile.center -| p),draw] (sa) {SA};
  \node[draw,right=of dv] (solvers) {Solvers};
  \draw[->] (annot) -- (bigotimes);
  \draw[->] (cfile.south east) -- (bigotimes);
  \draw[->] (bigotimes) -- (dv);
  \draw[->] (cfile) -- (sa);
  \draw[->] (sa) -- (p);
  \draw[->] (dv) -- (solvers);
\end{tikzpicture}
\vspace{1eX}
\caption{Verification using memory partitioning inferred by pointer analysis}
\label{fig:flow}
\end{center}
\vspace{-6eX}
\end{figure}
%
Our contribution targets this specific cooperation of SA and DV methods in the context of first-order logic solvers.
The verification process we propose is summarized by the flow diagram
in Figure~\ref{fig:flow}. The code to be verified is first given to the static analyzer to produce state invariants including a sound partitioning P of the program's memory. The partitioning P is exploited by a functor M which produces a memory model environment MME used by the DV tool to generate verification conditions into a logic theory supported by automatic solvers.
Our first contribution is the formalization of the functor M and of the information it needs from the static analysis. 
Secondly, we demonstrate that several existing pointer analyses may be used in this general framework. 
Thirdly, we implemented this functor in the \framac\ platform~\cite{fac/KirchnerKPSY15} between the plug-ins \Value\ for static analysis and \WP\  for deductive verification.
Finally, we propose a new pointer analysis exploiting a value analysis based on  abstract interpretation; this analysis is able to produce the memory model that reduces the verification effort of a relevant benchmark. 

\section{A Motivating Example}
\label{sec:over}

\begin{figure}[t]
\begin{center}
\begin{lstlisting}[style=customc,multicols=2]
typedef int32_t data_t;
typedef uint8_t pos_t;
typedef struct {
  data_t *in1, *in2, *in3, *in4;
  data_t *out1,*out2,*out3,*out4;
  pos_t  *pos1,*pos2,*pos3,*pos4;
               } intf4_t;
/*@ requires: 
 *   sep({args->in1,...,args->in4}, 
 *       args->out1,...,args->out4,  
 *       args->pos1,...,args->pos4); 
 *  ensures: 
 *   sorted_vals(&(args->out1),4);
 *  ensures: 
 *   perm(&(args->in1),&(args->out1),
 *        &(args->pos1),4); */
void sort4(intf4_t *args) {
  data_t **inArr   = 
    (data_t **) &(args->in1);
  data_t **outArr  = 
    (data_t **) &(args->out1);
  pos_t **posArr = 
    (pos_t **) &(args->pos1);
  /** init arrays from inputs */
  int32_t sortArr[4]; // values
  uint8_t permArr[4]; // permutation
  /*@ loop invariant: ... */
  for (int i = 0; i < 4; i++) { 
    sortArr[i] = *inArr[i];
    permArr[i] = i;
  }

  /* sorting algorithm on sortArr
   * with permutation in permArr */

  /** copy results to outputs */
  /*@ loop invariant: ... */
  for (int i = 0; i < 4; i++) {
    (*outArr[i]) = sortArr[i];
    (*posArr[i]) = permArr[i];
  }
}
\end{lstlisting} 
\vspace{-2eX}
\caption{Sorting function for $\ersigs=4$ inputs and outputs}
\label{lst:sort4}
\vspace{-4eX}
\end{center}
\end{figure}

We overview the issues targeted and the solution proposed in this work using the C code given in Figure~\ref{lst:sort4}.
This code is extracted from the C code generated by the compiler of a high level data flow language.
It combines at least three complex features of pointers in C.

The first feature is the duality of records and arrays, which is used here
to interpret the (large) list of arguments for a function as individual fields in a compound (record) type or as cells of an array.
Thus, the read of the $k$-th field ($k\ge 0$) named \code{fk} of a record stored at location \code{s} and using only fields of type $\tau$ may be written \code{s->fk} or \code{*(&(s->f0)+k)}, where \code{f0} is the first field.
%
It is debatable whether the C standard actually permits this form of dual indexing between records with fields of the same type and arrays~\cite{Stackoverflow-Q51737910}, but some programs, including this one, use this feature with success.
In our example, this duality is used in function \code{sort4} to ease the extraction of numerical values from the inputs and the storage of the sorted values in the outputs.
This first feature makes our running example more challenging, but the technique we propose is also effective when the parameters are encapsulated in arrays of pointers, \eg, when inputs and outputs are declared as a field of type array by \code{data_t* in[4]}.
The second feature is precisely the usage of arrays of pointers which is notoriously difficult to be dealt precisely by pointer analyses.
The third feature is the complex separation constraints between pointers stored in arrays, which leads to a quadratic number of constraints on the size of the array and complicates the task of DV tools.
In the following, we discuss in detail these issues and our approach to deal with them.

Inputs and outputs of \code{sort4} have the same type, \code{data_t}, which shall encapsulate a numerical value to be sorted. For simplicity, we consider only one field of \code{int32_t} type for \code{data_t}.
Type \code{pos_t} models an element of the permutation and denotes the destination position (an unsigned integer) of the value sorted.
The parameters of \code{sort4} are collected by type \code{intf4_t}: four pointers to \code{data_t} for input values, four pointers to \code{data_t} for output values, and four pointers to \code{pos_t} for the new positions of input values.

The function is annotated with pre/post conditions and with loop invariants.
The pre-condition requires (predicate \code{sep})
that 
(1) all pointers in \code{*args} are valid, i.e. point to valid memory locations, 
(2) the pointers in fields \code{in} are disjoint from any pointer in fields \code{out} and \code{pos}, and
(3) pointers in fields \code{out} and \code{pos} are pairwise disjoint.
Notice that the \code{in} fields may alias.
The post-condition states that  the
values pointed by the fields \code{out} are sorted (predicate \code{sorted_vals})
and, for each output $i$, the value of this output is equal to the value of the input $j$ such that \code{pos}[$j$] is $i$ 
(predicate \code{perm}).

The separation pre-condition is necessary for the proof of the post-condition because any aliasing between fields \code{out} may crush the results of the sorting algorithm. 
The encoding of this pre-condition in FOL is done by a conjunction of dis-equalities which is quadratic on the number of pointers concerned. More precisely, for $n$ inputs (and so $n$ outputs and $n$ positions), there are $O(n^2)$ such constraints. (In SL, this requirement is encoded in linear formulas.) The original code from which our example is inspired instantiate $n$ with $24$ and therefore generates a huge number of dis-equalities.
Several techniques have been proposed to reduce the number of dis-equalities generated by the separation constraints.
For example, a classic technique is assigning a distinct logic value (a color) to each pointer in the separated set. This technique does not apply in our example if the type \code{data_t} is a record with more than one field because the color shall concern only the numerical value to be sorted.

As an alternative, we propose to use precise points-to analyses to lift out such constraints and to simplify the memory model used for the proof of the function.
Importantly, we perform a per-call proof of \code{sort4}, instead of a unitary proof.
For each call of \code{sort4}, the static analysis tries to check
that the separation pre-condition is satisfied and provides a model for the memory where the pointers are dispatched over disjoint zones.
Unfortunately, the precision of the points-to analyses (and consequently the number of separation constraints discharged) may change radically with the kind of initialization done for the arguments of \code{sort4}.
We will illustrate this behavior for two calls of \code{sort4} given in Figure~\ref{lst:main}: the call in listing (a) uses variables and the one in listing (b) uses arrays.
Notice that each call satisfies the separation pre-condition of \code{sort4}.

\begin{figure}[t]
\begin{center}
\begin{minipage}{.45\textwidth}
\begin{lstlisting}[style=customc,caption=(a) using variables]{} %{src/main-var.c} 
  data_t  df_1,df_2,...,df_8;
  pos_t   pf_1,pf_2,pf_3,pf_4;
  intf4_t SORT = {
    .in1=&df1,  .in2=&df2,
    .in3=&df3,  .in4=&df4,
    .out1=&df5, .out2=&df6,
    .out3=&df7, .out4=&df8,
    .pos1=&pf1, .pos2=&pf2,
    .pos3=&pf3, .pos4=&pf4 };

  df_1 = nondet_data();
  df_2 = nondet_data();
  df_3 = nondet_data();
  df_4 = nondet_data();

  sort4(&SORT);
\end{lstlisting}
%
%
%
\end{minipage}\hfill
\begin{minipage}{.45\textwidth}
\begin{lstlisting}[style=customc,caption=(b) using arrays]{} %{src/main-arr.c} 
  data_t  df[8];
  pos_t   pf[4];
  intf4_t SORT = {
    .in1=df+1,  .in2=df+2,
    .in3=df+3,  .in4=df+4,
    .out1=df+5, .out2=df+6,
    .out3=df+7, .out4=df,
    .pos1=pf,   .pos2=pf+1,
    .pos3=pf+2, .pos4=pf+3 };

  df[1] = nondet_data();
  df[2] = nondet_data();
  df[3] = nondet_data();
  df[4] = nondet_data();

  sort4(&SORT);
\end{lstlisting}
%
%
%
\end{minipage}
\vspace{-2eX}
\caption{Two calls for the sorting function using different initialization}
\label{lst:main}
\vspace{-6eX}
\end{center}
\end{figure}

\mypar{Typed memory model:} 
For completeness, we quickly present first how DV tools using FOL deal with our example using the Burstall-Bornat model. 
In this model, the memory is represented by a set of array variables, each array corresponding to a (pre-defined, basic) type of memory locations. 
For our example, the memory model includes six array variables: 
\code{M_int32}, \code{M_uint8}, \code{M_int32_ref}, \code{M_uint8_ref}, \code{M_int32_ref_ref}, \code{M_uint8_ref_ref} storing values of type respectively \code{int32_t}, \code{uint8_t}, \code{int32_t*}, \code{uint8_t*}, \code{int32_t**} and \code{uint8_t**}.
Program variables are used as indices in these arrays, \eg, variable \code{inArr} is an index in array \code{M_int32_ref_ref} and \code{sortArr} is index of \code{M_int32}. 

The separation pre-condition of \code{sort4} is encoded by dis-equalities, e.g., \code{M_int32_ref[args_in4] <> M_int32_ref[args_out1]} where \code{args_in4} is bound to the term $\textit{shift}(\code{M_int32_ref_ref[args], in4})$ which encodes the access to the memory location \code{&(args->in4)} using the logic function \code{shift};
\code{args_out1} is defined similarly.
However, these dis-equalities are not propagated through the assignments at lines 18--23 in Figure~\ref{lst:sort4}, which interpret the sequence of (input/output/position) fields as arrays.
Therefore, additional annotations are required to prove the correct initialization of the output at lines 39--41.
Some of these annotations may be avoided using our method that employs pointer analyses to infer precise memory models, as we show below.

\mypar{Base-offset pointer analysis:}
Consider now a pointer analysis which is field and context sensitive, and which computes  
an over-approximation of the value of each pointer 
expression at each program statement.
%
The over-approximation, that we name \emph{abstract location}, is built upon the standard concrete memory model of~C~\cite{jar/LeroyB08}. 
An abstract location is \emph{a partial map} between the set of program's variables and the set of intervals in $\NN$. 
An element of this abstraction, $(\cvar,\mathabs{i})$, 
denotes the symbolic (\ie, not related with the locations in the virtual memory space used during the concrete execution) memory block that starts at the location of the program variable $\cvar$ (called also \emph{base}),
and the abstraction by an interval $\mathabs{i}$ of the set of possible offsets (in bytes) inside the symbolic block of $\cvar$ to which the pointer expression may be evaluated. 
In this memory model, symbolic blocks of different program variables are implicitly separated: it is impossible to move from the block of one variable to another using pointer arithmetic. 
The memory model is modeled by a set of logic arrays, one for each symbolic block. The over-approximation computed by the analysis allows to dispatch a pointer expression used in a statement on these arrays.

In our example, for the call of \code{sort4} in Figure~\ref{lst:main} (a), the memory model includes the symbolic blocks for program's variable \code{df}$i$, \code{pf}$i$ and \code{SORT}.
The above analysis computes for 
the pointer expressions \code{args->in1} and \code{*(args->in1)}
at the start of \code{sort4}, the abstract location 
$\{(\code{SORT},[0,0])\}$ and\ $(\code{df1},[0,0])$ respectively.
The abstract locations for the pointer expressions involving other fields of \code{args} are computed similarly. 
The separation pre-condition of \code{sort4} is implied by these abstract locations.
After the fields of \code{args} are interpreted as arrays (lines 18--23 of \code{sort4}), the pointer expression \code{outArr+i} at line 39, 
where \code{i} is restricted to the interval $[0,3]$, 
is over-approximated to the abstract location\todocomment{MS: keep simple, see comments. BY: I agree}{}
$\{(\code{SORT},[16,31])\}$. 
Similarly, \code{inArr+i} is abstracted by $\{(\code{SORT},[0,15])\}$.
Therefore, the left value given by the pointer expression 
\code{outArr[i]} (at line 39) is (precisely) computed to be  
$\{(\code{df5},[0,0]),...,(\code{df8},[0,0])\}$.
This allows proving the correctness of the output computed by \code{sort4}.

For the call in Figure~\ref{lst:main}~(b), the memory model includes symbolic blocks for program's variable \code{df}, \code{pf} and \code{SORT}.
The analysis computes for pointer expressions \code{args->in1} and \code{*(args->in1)} (used at the start of \code{sort4}), 
the abstract location 
$\{(\code{SORT},[0,0])\}$ \resp\ $(\code{df},[0,3])$,
which also allows to prove the separation pre-condition.
The interpretation of fields as arrays (lines 18--23) leads to the abstract location $\{(\code{df},[1,4])\}$ for \code{inArr+i}, which is very precise. 
However, because the initialization of the field \code{SORT.out4} at line 18 in Figure~\ref{lst:main}~(b) breaks the uniformity of the interval, the pointer expression \code{outArr+i} (at line 39) is over-approximated to $\{(\code{df},[0,7])\}$. This prevents the proof of the post-condition.

In conclusion, such an analysis is able to infer a sound memory model that offers a finer grain of separation than the typed memory model. However, it is not precise enough to deal with the array of pointers and field duality in records.

\mypar{Partitioning analysis:}
Based on the base-offset pointer analysis above, we define in Section~\ref{ssec:slice} a new analysis that computes for each pointer expression an abstract location that collects a finite set of \emph{slices of symbolic blocks}, \ie, 
the abstraction is a partial mapping from program's variables to sets of intervals representing offsets in the block.
With this analysis, the abstract location computed for \code{outArr+i} (at line 39 of \code{sort4}, call in Figure~\ref{lst:main} (b)) is more precise, \ie, $\{\code{df}\mapsto\{[5,7], [0,0]\}\}$,
and it allows to prove the post-condition for \code{sort4}. 
Notice that the analysis computes \emph{a finite set of slices} in symbolic blocks whose concretizations (sets of locations) are pairwise disjoint.
For this reason, this analysis may be imprecise if its parameter fixing the maximum size of this set is exceeded.
This analysis also deals precisely with the call of \code{sort4} in Figure~\ref{lst:main} (a).

\mypar{Dealing with different analyses:}
The above comments demonstrate the diversity of results obtained for the memory models for different points-to analysis algorithms. One of our contributions is to define a generic interface for the definition of the memory model for the DV based on the results obtained by static analyses doing points-to analysis (SPA). 
This interface eases the integration of a new SPA algorithm and the comparison of results obtained with different SPA algorithms. 
We formalize this interface in Section~\ref{sec:pbmm} and instantiate it for different SPA algorithms in Section~\ref{sec:mm-with-ai}. Our results are presented
in Section~\ref{sec:experiments}. 




\section{Generating Verification Conditions}

To fix ideas, we recall the basic principles of generating verification conditions (VC) using a memory model by means of a simple C-like language.

\subsection{A \clight Fragment}

We consider a fragment of \clight~\cite{jar/BlazyL09} that excludes casts, union types and multi-dimensional arrays. We also restrict the numerical expressions to integer expressions. The syntax of expressions, types and atomic statements is defined by the grammar in Figure~\ref{fig:syntax}.
This fragment is able to encode all assignment statements in Figures~\ref{lst:sort4}--\ref{lst:main} using classic syntax sugar (\eg, \code{**(arr + i)} for \code{*arr[i]}, \code{&((*args).in1)} for \code{&(args->in1)}).
Complex control statements can be encoded using the standard way. 
%
User defined types are pointer types, static size array types,
and record types. A record type declares a list of typed fields with
names from a set $\cflds$; for simplicity, we suppose that each field
has a unique name. 
We split expressions into integer expressions and address expressions to ease their typing. Expressions are statically typed by a type $\cty$ in $\ctyps$. When this information is needed, we write $\cexp^\cty$.

\begin{figure}[t] 
$\begin{array}{ll p{5mm} ll}
\csyn{n} \in \mathbb{N}, \csyn{k} \in \mathbb{Z} & \mbox{integer constants}
& &
\cinttyps & \mbox{integer type in }\{\mathtt{i8},\mathtt{u8},\mathtt{i16},\ldots,\mathtt{u64}\}
\\
\crec\in\crecs & \mbox{record type names} 
& & 
\cfld\in\cflds & \mbox{field names}
\\
\cvar \in \cvars & \mbox{program variables}    
& & 
\cop \in \cops & \mbox{unary and binary arithmetic operators}
\\[3mm]
\end{array}$

$\begin{array}{lrll}
\mbox{scalar types }   & \cstyps \ni \csty & ::=
   \cinttyps \mid 
   \cptr{\cty} 
\\[1mm]   
\mbox{program types } & \ctyps \ni \cty & ::=  
   \csty \mid 
   \crec \mid 
   \carr{\csty}{\csyn{n}}
\\[1mm] 
\mbox{expressions } & \cexps \ni \cexp & ::=
    \ciexp \mid 
    \caexp 
\\[1mm]
\mbox{integer expressions } & \ciexps \ni \ciexp & ::=
	\csyn{k} \mid 
	\clval \mid
	\cop~ \ciexp \mid 
	\ciexp ~\cop~ \ciexp'
\\[1mm]
\mbox{address expressions } & \caexps \ni \caexp & ::=
    \cnull \mid
    \clval \mid 
    \csyn{\&}\clval \mid
    \caexp + \ciexp
\\[1mm]
\mbox{left-values } & \clvals \ni \clval  & ::= 
	\cvar \mid
        \clval\csyn{.}\cfld \mid
	\csyn{*} \caexp
\\[1mm]
\mbox{statements } & \cstmts \ni \cstmt & ::=
    \clval \csyn{=} \cexp \mid 
    \cassert~\cexp 
\end{array}$
\caption{Syntax of our \clight fragment}
\label{fig:syntax}
\vspace{-2eX}
\end{figure}

%

We choose to present our work on this simple fragment for readability. However, our framework may be extended to other constructs.
For example, our running example contains struct initialization. Struct assignment may be added by explicit assignment of fields.
Type casting for arithmetic and compatible pointer types (\ie, aligned on the same type) may be dealt soundly in DV tools employing array-based memory models using the technique in~\cite{RakamaricH09}.
Functions calls may be also introduced if we choose context-sensitive SA. 
In general, DV tools conduct unit proofs for functions. We restrict this work to whole-program proofs, because it avoids the requirement that SA is able to conduct analyses starting with function's pre-conditions.

\subsection{Memory Model}
\label{ssec:amm}

We define the denotational semantics of our language using an environment called \emph{abstract memory model} (AMM).
(This name is reminiscent of the first abstract memory model defined in~\cite{inria/LeroyAB12,jar/LeroyB08} for \textsf{CompCert}. We enriched it with some notations to increase readability of our presentation.)
Figure~\ref{fig:am} summarizes the elements of this abstract memory model.
The link between the abstract memory model and the concrete standard memory model is provided in Appendix~\ref{sec:sem}.


\begin{figure}[t]
\vspace{-2eX}
\fbox{
$\begin{array}[t]{rrl}
\multicolumn{3}{l}{\mathbf{sig}~\textit{AMM}\,:} \\ 
~~
\mathbf{type}
& \multicolumn{2}{l}{\amloc \triangleq \cvars \times \NN} \\ 
\mathbf{ops}
& \ambase : & \cvars \ft \amloc  \\
& \amshft : & \amloc \ft \NN \ft \amloc
\end{array}
\begin{array}[t]{rrl}
~\\
\mathbf{types} 
		& \ammem, 
		& ~~\amval \triangleq \amvint(\ZZ) \mid \amvptr(\amloc) \\ 
\mathbf{ops} 
& \amload  : & \ammem \ft \cstyps \ft \amloc \ft \amval_\bot \\
& \amstore : & \ammem \ft \cstyps \ft \amloc \ft \amval \ft \ammem_\bot 
\end{array}$}
\caption{Abstract signature for the concrete memory model}
\label{fig:am}
\vspace{-4eX}
\end{figure}

\comment{
The state of the memory is represented by an abstract data type $\ammem$.
A memory state is a collection of \emph{memory blocks}.
A block denotes a slice of memory storing a sequence of values of type $\amval$. Blocks are pairwise disjoint.
Pointer values, called locations, denote pairs $(b,o)$ composed of 
	a block 
	$b$ and a byte offset $o$ ($\ge 0$) inside the block.
We denote by $\amloc$ the set of such pairs and provide two operations to build them:
}
The states of the memory are represented by an abstract data type $\ammem$
which associates locations of type $\amloc$ to values in the type $\amval$.
Locations are pairs $(b,o)$ where 
$b$ is the identifier of a symbolic block and 
$o$ is an integer giving the offset of the location in the symbolic block of $b$.
Because we are not considering dynamic allocation, symbolic blocks are all labeled by program's variables. Thus we simplify the concrete memory model by replacing block identifiers by program variables.
Values of type $\amloc$ are built by two operations of $\textit{AMM}$: 
$\ambase(\cvar)$ gives the location of a program variable $\cvar$
and $\amshft(\ell,n)$ computes the location obtained by shifting the offset of location $\ell$ by $n$ bytes.
The shift operation abstracts pointer arithmetics.
%
The typing function $\ctyof{.}$ is extended to elements of 
$\amloc$ based on the typing of expressions used to access them.
%
Some operations are partial and we denote by $\bot$ the undefined value. A set $A$ extended with the undefined value is denoted by $A_\bot$. 
The axiomatization of loading and storing operations is similar to the one in~\cite{inria/LeroyAB12,jar/LeroyB08}.


\subsection{Semantics}

Figure~\ref{fig:sem} defines the rules of the semantics
using the abstract memory model, via the overloaded functions $\sem{\cdot}$.
The semantic functions are partial: the undefined case $\bot$ cuts the evaluation.
The operators $\widehat{\cop}$ are interpretations of operations $\cop$ over integer types $\cinttyps$.
The functions $\cfoff{\cdot}$ and $\ctsz{\cdot}$ are defined by the Application Binary Interface (ABI) and depend on the architecture.
Conversions between integer values are done using function $\ctyto{\cdot}{\cdot}$.

\begin{figure}[ht]
\begin{minipage}[t]{0.65\textwidth}
$\begin{array}{ll}
\sem{\cdot}:\cstmts \ft \ammem & \ft \ammem_\bot
\\\hline
\sem{\clval^{\csty} \csyn{=} \cexp}(\amv)
    & \triangleq \amstore(\amv, \csty, \sem{\clval}(\amv), \sem{\cexp}(\amv)) 
\\
\sem{\csyn{\cassert~\cexp}}(\amv) 
    & \triangleq \mbox{if } \sem{\cexp}(\amv) \not\sim 0 \mbox{ then } \amv \mbox{ else } \bot 
\end{array}$
\end{minipage}
\begin{minipage}[t]{0.3\textwidth}
$\begin{array}{ll}
\sem{\cdot}:\cexps & \ft \ammem \ft \amval_\bot
\\\hline
\sem{\ciexp}(\amv)
    & \triangleq \amvint(\sem{\ciexp}(\amv))
\\
\sem{\caexp}(\amv)
    & \triangleq \amvptr(\sem{\caexp}(\amv))
\end{array}$
\end{minipage}

\vspace{1eX}
\begin{minipage}[t]{0.58\textwidth}
$\begin{array}[t]{ll}
\sem{\cdot}:\ciexps&\ft\ammem\ft\ZZ_\bot
\\\hline
\sem{\csyn{i}}(\amv)
    & \triangleq \csyn{i} \\
\sem{\clval^{\carith}}(\amv)
    & \triangleq i, 
\amvint(i) =\amload(\amv, \carith, \sem{\clval}(\amv)) \\
\sem{\csyn{\cop~\ciexp}}(\amv)
    & \triangleq \widehat{\cop}(\sem{\ciexp}(\amv)) \\
\end{array}$
\end{minipage}
\begin{minipage}[t]{0.36\textwidth}
$\begin{array}[t]{ll}
\sem{\cdot}:\clvals&\ft\ammem\ft\amloc_\bot
\\\hline
\sem{\cvar}(\amv)
	& \triangleq \ambase(\cvar) \\
\sem{\clval\csyn{.}\cfld}(\amv)
	& \triangleq \amshft(\sem{\clval}(\amv),\cfoff{\cfld}) \\
\sem{\csyn{*}\caexp}(\amv)
	& \triangleq \sem{\caexp}(\amv)
\end{array}$
\end{minipage}

\vspace{1eX}
$\begin{array}{ll}
\sem{\cdot}:\caexps\ft\ammem&\ft\amloc_\bot
\\\hline
\sem{\cnull}(\amv)
    & \triangleq \ambase(\cnull) \\
\sem{\clval^{\carr{\csty}{n}}}(\amv)
    & \triangleq \sem{\clval}(\amv) \\
\sem{\clval^{\cptr{\cty}}}(\amv)
    & \triangleq \ell \mbox{ where } \amload(\amv,\cptr{\cty},\sem{\clval}(\amv))=\amvptr(\ell) \\
\sem{\csyn{\& \clval}}(\amv)
    & \triangleq \sem{\clval}(\amv) \\
\sem{\csyn{\caexp^{\cptr{\cty}}+\ciexp}}(\amv) 
    & \triangleq \amshft(\sem{\caexp}(\amv),
        \ctsz{\cty}
        \times\ctyto{\sem{\ciexp}(\amv)}{\mathtt{u32}})
\end{array}$
\caption{Semantics of our \clight\ fragment}
\label{fig:sem}
\vspace{-4eX}
\end{figure}


\subsection{Generating Verification Conditions}

Verification conditions (VC) are generated from Hoare's triple 
	$\{P\}~\cstmt~\{Q\}$
with $P$ and $Q$ formulas in some logic theory $\llogics$ used
for program annotations and $\cstmt$ a program statement. 
The classic method \cite{ipl/Leino05,Flanagan/2001} is built on 
the computation of a formula 
$R_\cstmt(\vec{v}_{b},\vec{v}_{e})$ in $\llogics$
specifying the relation between the states of the program before
and after the execution of $\cstmt$, which are represented by the
set of logic variables $\vec{v}_{b}$ resp. $\vec{v}_{e}$. 
The VC built for the above Hoare's triple is 
$\forall\vec{v}_{b},\vec{v}_{e}.~\big(P(\vec{v}_{b})\land R(\vec{v}_{b},\vec{v}_{e})\big)\implies Q(\vec{v}_{e})$
and it is given to solvers for $\llogics$ to check its validity.
In the following, we denote by $\lterms$ the set of logic terms built in the logic theory $\llogics$
using the constants, operations, and variables in a set $\lvars$.
For a logic sort $\tau$, we designate by $\lterms_{\tau}$ the terms
of type $\tau$.

\mypar{Compilation environment:}
Formula $R_\cstmt(\cdot,\cdot)$ is defined based on the dynamic semantics of statements, like the one given in Figure~\ref{fig:sem} for our language.
The compilation of this semantics into formulas $\llogics$ uses a \emph{memory model environment} (called simply environment) that implements the interface of the abstract memory model given in Figure~\ref{fig:am}.
This environment changes at each context call and keeps the information required by the practical compilation into formulas, \eg, the set of variables used for modeling the state at the current control point of this specific context call.
Figure~\ref{fig:MME} defines the signature of memory environments.  


\begin{figure}[b] 
\vspace{-4eX}
\fbox{$
\\[1mm]
\begin{array}[t]{rl}
\multicolumn{2}{l}{\mathbf{sig}~\mathsf{MME}\,:} \\
~~ 
\mathbf{type} & \mmloc \\ 
\mathbf{ops} & \mmbase :  \cvars \ft \mmloc \\
			 & \mmshft :  \mmmem \ft \mmloc \ft \lterms_\lintegers \ft \mmloc_\bot 
\end{array}
\begin{array}[t]{rl}
~\\
\mathbf{types} 
	& \mmmem,\ \mmval \triangleq \mmvint(\lterms_\lintegers) \mid \mmvptr(\mmloc)  \\ 
\mathbf{ops} & \mmload : \mmmem \ft \cstyps \ft \mmloc \ft \mmval_\bot \\ 
		     & \mmstore :  \mmmem \ft \cstyps \ft \mmloc \ft \mmval \ft
               (\mmmem \times \lterms_\lbools)_\bot
\end{array}$}
\caption{Signature of the memory model environments}
\label{fig:MME}
\vspace{-4eX}
\end{figure}

The types $\mmmem$ and $\mmloc$ encapsulate 
information about the program states and memory locations respectively.
Notice that the logical representation of locations is hidden by this interface, which allows to capture very different memory models.
The compilation information about the values stored is given by the type $\mmval$,
which represent integers by integer terms in $\ltheory$, \ie, in the set $\lterms_\lintegers$.
Operation $\mmshft$ implements arithmetics on locations by an integer term.
Operation $\mmstore$ encapsulates the updating of the environment by an assignment and produces a new environment and a term in $\lterms_\lbools$, i.e., a formula of $\ltheory$.%

\mypar{Prerequisites on the logic theory:}
For DV tools based on first order logic, the theory $\llogics$ is a multi-sorted FOL that embeds 
the logic theory used to annotate programs (which usually includes boolean and integer arithmetics theories)
and the McCarthy's array theory~\cite{ifip/McCarthy62} employed by the Burstall-Bornat memory model~\cite{Bornat00} to represent atomic memory blocks.
The memory model environment associates to each memory blocks a set of logic array variables using $\mmbase$ operations.
It encodes 
the operations $\mmload(\mmv,\cty,\ell)$ \resp\ $\mmstore(\mmv,\cty,\ell,v)$ 
into 
logic array operations $\textit{read}(a,o)$ \resp\ $\textit{store}(a,o,v)$,
where $a$ is the array variable for the symbolic block $b$ of location $\ell$ that stores values of type $\cty$ and $o$ is the offset of $\ell$ in $b$.
$\llogics$ also embeds abstract data types (or at least polymorphic pairs with component selection by $\fst$ and $\snd$), and uninterpreted functions. 
Polymorphic conditional expression 
``$(e_{\textit{cond}})?e_{\textit{true}}:e_{\textit{false}}$'' 
are also needed.

In the following, we use the logic theory above $\llogics$ and
suppose that an infinite number of fresh variables can be generated.
To ease the reading of environment definitions, we distinguish the
logic terms by using the mathematical style and by underlining the
terms of $\llogics$, \eg, $\logic{x+x}$. For example, the logic
term $\logic{\textit{read}(}\vcgen{\mmv(b)}\logic{,4+x})$
is built from a VC generator
term $\vcgen{\mmv(b)}$ that computes a logic term of array type
and the logic sub-term $\logic{\textit{read}(\cdot,4+x)}$.

\mypar{Example:}
Consider the Hoare's triple $\{P\}~\csyn{(*(\&r.f))^{i8}=5}~\{Q\}$.
Let $\mml_0$ be $\mmshft(\mmv_0,\mmbase(\csyn{r}),\cfoff{\csyn{f}})$,
where $\mmv_0$ (\resp\ $\mmv_1$) is the environment for the source state (\resp\ modified by the store for the destination state);
that is
$\mmv_1, \phi_1  \triangleq   \mmstore(\mmv_0, \csyn{i8}, \mml_0,
                                 \mmvint(5))$.
The formula $\logic{P}$ (\resp\ $\logic{Q}$) is generated from $P$ (\resp\ $Q$) using compilation environment $\mmv_0$ (\resp\ $\mmv_1$).
Then the VC generated by the above method is 
$\logic{P}\land \phi_{1}\implies \logic{Q}$.
Notice that the above calls of the environment's operations follow the order given by the semantics in Figure~\ref{fig:sem}, except for the failure cases.
Indeed, to simplify our presentation, we consider that statement's pre-condition includes the constraints that eliminate runs leading to undefined behaviors.
Therefore, the VC generation focuses on encoding in $R_\cstmt(\cdot,\cdot)$ the correct executions of statements.



{
\abovedisplayskip=8pt plus 3pt minus 7pt
\belowdisplayskip=8pt plus 3pt minus 7pt

\section{Partition-based Memory Model}
\label{sec:pbmm}

We define a functor that produces memory models environments implementing the interface on Figure~\ref{fig:MME} from the information inferred by a pointer analysis. 
The main idea is that the SA produces a finite partitioning of symbolic blocks into a set of pairwise disjoint sub-blocks and each sub-block is mapped to a specific set of array logic variables by the compilation environment. 
We first formalize the pre-requisites for the pointer analysis using a signature constrained by well-formed properties.
Then, we define the functor by providing an implementation for each element of the interface on Figure~\ref{fig:MME}.

\subsection{Pointer Analysis Signature} 

A necessary condition on the pointer analysis is its soundness.
To ease the reasoning about this property of analysis, we adopt the abstract interpretation~\cite{popl/CousotC77} framework.
In this setting, a SA computes an abstract representation $\mathabs{s}$ of the set of concrete states reached by the program's executions before the execution of each statement.
The abstract states $\mathabs{s}$ belong to a complete lattice $(\mathabs{S},\mathabs{\sqsubseteq})$ which is related to the set of concrete program configurations $\mathit{State}$ by a pair of functions 
$\alpha:2^\mathit{State}\ft\mathabs{S}$ (abstraction) and $\gamma:\mathabs{S}\ft 2^\mathit{State}$ (concretization) forming a Galois connection. 
In the following, we overload the symbol $\gamma$ to denote concretization functions for other abstract objects.

\begin{figure}[t]
\fbox{$
\begin{array}[t]{rl}
\mathbf{sig}~\mathsf{PA}\,: \\
\mathbf{type}~ 
	& \mmvdloc \\
\mathbf{ops}~ 
	& \mmvdfont{base} : \cvars \to \mmvdloc \\
	& \mmvdfont{domain} : \mmvdloc \ft \pset{\mmvdblk} 
\end{array}\quad
\begin{array}[t]{rl}
~\\
\mathbf{type}~
	& \mmvdst \\
\mathbf{ops}~
	& \mmvdfont{load} : \mmvdst \to \cptrtyps \to \mmvdloc \to \mmvdloc \\
	& \mmvdfont{shift} : \mmvdst \to \mmvdloc \to \lterms_\lintegers \to \mmvdloc 
\end{array}\quad
\begin{array}[t]{rl}
~\\
\mathbf{type}~
	& \mmvdblk \\
\mathbf{ops}~ 
	& \mmvdfont{base} : \mmvdblk \ft \cvars
\\
	& \mmvdfont{slice} : \mmvdblk \to \lterms_\lintegers \to \lterms_\lbools
\end{array}	
$}
\begin{eqnarray}
\textrm{disjointness:} & & 
	\forall \mathabs{b}_1,\mathabs{b}_2\in\mmvdblk
	\cdot \mathabs{b}_1\neq\mathabs{b}_2 \Rightarrow 
			\gamma(\mathabs{b}_1) \cap \gamma(\mathabs{b}_2) = \emptyset 
\label{eq:bdisjoin}
\\
\textrm{completeness:} & & 
	\forall \cvar\in\cvars\ \forall i\in[0,\csizeof(\ctyof{\cvar})-1)\ 
		\exists\mathabs{b}\in\mmvdblk \cdot 
			(\cvar,i)\in\gamma(\mathabs{b})
\label{eq:bcomplete}
\\
\textrm{unique base:} & & 
	\forall \mathabs{b}\in\mmvdblk\ \exists! \cvar\in\cvars \cdot 
		\gamma(\mathabs{b}) \subset \{ (\cvar,i) \mid i \in \NN \} 
\label{eq:bbase}
\\
\textrm{sound }\mmvdblk\textrm{ ops:} & & 
	\forall\mathabs{b}\in\mmvdblk\cdot
		\gamma(\mathabs{b})=\{ (\cvar,i) \in \amloc \mid 
			\cvar = \mmvdfont{base}(\mathabs{b}) \land 
			\mmvdfont{slice}(\mathabs{b},\logic{i})=\textit{true}
	\}~~~~
\label{eq:bbaseslice}
\\
\textrm{sound }\mmvdloc\textrm{ ops:}	
	& & \forall\mathabs{\ell}\in\mmvdloc\ 
			\forall\ell\in\gamma(\mathabs{\ell})\ 
			\exists\mathabs{b}\in\mmvdfont{domain}(\mathabs{\ell})\cdot 
				\ell\in\gamma(\mathabs{b})
\label{eq:bdomain}
\\
\textrm{sound }\mmvdst\textrm{ ops:}	
	& &
	\forall\cstmt\ \forall\mathabs{s}\in\mmvdst(\cstmt)\ 
		\forall\mathabs{\ell}\in\mathabs{s}\cdot \nonumber \\
	& & \quad\quad\gamma(\mmvdfont{shift}(\mathabs{s},\mathabs{\ell},e)) 
		\supseteq \{  \amshft(\ell,i) \mid \ell\in\gamma(\mathabs{\ell}), 
		\amv\in\gamma(\mathabs{s}), i\in\sem{e}(m) \}~~~~
\label{eq:bshift}
\\
	& &
	\forall\cstmt\ \forall\mathabs{s}\in\mmvdst(\cstmt)\ 
		\forall\mathabs{\ell}\in\mathabs{s}\cdot \nonumber \\
	& & \quad\quad\gamma(\mmvdfont{load}(\mathabs{s},\cptr{\cty},\mathabs{\ell})) \supseteq
		\{ \amload(\amv,\cptr{\cty},\ell) \mid \ell\in\gamma(\mathabs{\ell}), 
		\amv\in\gamma(\mathabs{s}) \}~~~~
%
\label{eq:bload}
\end{eqnarray}
\vspace{-4eX}
\caption{A signature for pointer analysis and its properties}
\label{figure:model-domain-sig}
\vspace{-4eX}
\end{figure}

Aside being sound, the SA shall be context sensitive and provide, for each context call, an implementation of the signature on Figure~\ref{figure:model-domain-sig}. 
The values of $\mmvdst$ provides, for each statement of the current context, the abstract state in $\mathabs{S}$ computed by the analysis.
%
%
The type $\mmvdloc$ represents the domain of abstract values computed for the pointer expressions in abstract states. 
The concretization function $\gamma: \mmvdloc \to \pset{\amloc}$ maps abstract locations to sets of concrete locations.

The type $\mmvdblk$ stands for the set of pairwise disjoint abstract blocks partitioning the symbolic memory blocs, for the fixed specific context call. 
The concretization function for abstract blocks 
$\gamma:\mmvdblk\to\pset{\amloc}$ maps blocks to set of concrete locations.
Equations (\ref{eq:bdisjoin}) and (\ref{eq:bcomplete}) in Figure~\ref{figure:model-domain-sig} specify that 
abstract blocks in $\mmvdblk$ shall form a partition of the set of concrete locations available in symbolic blocks such that an abstract block belongs to a unique symbolic block.

The operation $\mmvdfont{base}(\mathabs{b})$ returns the symbolic block to which $\mathabs{b}$ belongs, represented by the program variable labeling this symbolic block.
The range of an abstract block $\mathabs{b}$ inside its symbolic block is specified by the operation $\mmvdfont{slice}(\mathabs{b},e)$, which returns a formula (boolean term in $\lterms_\lbools$) that constrains $e$ to be in this range. 
The soundness of the $\mmvdfont{base}$ and $\mmvdfont{slice}$ operations is specified by equation (\ref{eq:bbaseslice}).
%
%
The set of abstract blocks covered by an abstract location is provided by the operation $\mmvdfont{domain}$, whose soundness is specified by equation (\ref{eq:bdomain}).
%
The operation $\mmvdfont{base}$ abstracts the offset $0$ of a program variable. Abstract locations may be shifted by an integer term using operation 
$\mmvdfont{shift}$. 
Operation $\mmvdfont{load}(s,\cptr{\cty},\mathabs{\ell})$ computes the abstract location stored at $\mathabs{\ell}$ in some context $s$, \ie, 
it dereferences $\mathabs{\ell}$ of type $\cptr{\cptr{\cty}}$ for some $\cty$. (We denote by $\cptrtyps$ the set of all pointer types in the program.) 
The last two operations shall be sound abstract transformers on abstract locations, as stated in equations (\ref{eq:bshift}) \resp\ (\ref{eq:bload}).



\vspace{-3eX}
\subsection{A Functor for Memory Model Environments}
\label{ssec:functor}

We define now our functor that uses the signature $\mathsf{PA}$ to define the elements of the memory model environment $\mathsf{MME}$ defined in Figure~\ref{fig:MME}.
To disambiguate symbols, we prefix names of types and operations by the name of the signature or logic theory when necessary.

\mypar{Environment's type:} 
A compilation environment $\mmv\in\mmmem$ stores the mapping to abstract states from $\mathsf{PA}$ and
and a total function that associates to each abstract block in $\mathsf{PA}.\mmvdblk$ a logic variable in $\lvars$:
\begin{equation}
\mathsf{MME}.\mmmem\triangleq 
\mathsf{PA}.\mmvdst \times [\mathsf{PA}.\mmvdblk\to \ltheory.\lvars] 
\end{equation}
where $[A\to B]$ denotes the set of total functions from $A$ to $B$, \ie, $B^A$.
We designate by $\mmv_s$ and $\mmv_\epsilon$ the first and second component of some $\mmv\in\mmmem$.

If an abstract block $\mathabs{b}$ stores only one type of values, 
the logic variable $\mmv_\epsilon(\mathabs{b})$ 
has type $\larray(\ZZ,\tau)$ where $\tau$ is the logic type for the values stored. 
For blocks storing integer values (\ie, $\carith$), $\tau$ is naturally (logical) $\ZZ$ or $\NN$. 
For blocks storing pointer values, $\tau$ is $\ZZ\times\ZZ$,
$(b,o)$ where the $b$ denotes the abstract block of the location 
and $o$ represents the location's offset. 
We denote by $\logic{\mathabs{b}}$ the integer constant that uniquely identifies $\mathabs{b}\in\mmvdblk$.
If an abstract block $\mathabs{b}$ stores values of both kinds of scalar types (notice that only scalar values are stored in array-based models), the logic variable $\mmv_\epsilon(\mathabs{b})$ has the type pair of arrays, 
$(\larray(\ZZ,\ZZ),\larray(\ZZ,\ZZ\times\ZZ))$ where the first array is used for integer values and the second one for pointer values.
For readability, we detail here only the case of homogeneously typed blocks.
Notice that the mapping $\mmv_\epsilon$ binds fresh array variable names to abstract blocks changed by $\mmstore$ operation.

\mypar{Locations' type:} 
The type $\mathsf{MME}.\mmloc$ collects the logic encoding of locations as a pair of integer terms $(e_b,e_o) \in \lterms_\lintegers \times \lterms_\lintegers$ together with the abstract location $\ell$ provided by the static analysis, \ie, 
$\mathsf{MME}.\mmloc\ \triangleq\ 
	\lterms_{\lintegers \times \lintegers} \times \mathsf{PA}.\mmvdloc
$. 
Intuitively, in the logic pair $(e_b,e_o)$, 
	$e_b$ is interpreted as an abstract block identifier and
	$e_o$ models the offset of the location in the \emph{symbolic block of the abstract block} $e_b$, \ie, an integer in the slice of $e_b$.

\mypar{Locations' operations:} 
The values of $\mathsf{MME}.\mmloc$ are built by two operations $\mathsf{MME}.\mmbase$ and $\mathsf{MME}.\mmshift$ defined as follows.
For a program variable $\cvar$, $\mathsf{MME}.\mmbase(\cvar)$ is based on the abstract location $\mathabs{\ell}$ returned by $\mathsf{PA}.\mmbase(\cvar)$. The domain of $\mathabs{\ell}$ shall have only one abstract block $\mathabs{b}$ because program variables are located at the start of symbolic blocks. 
Moreover, the term denoting the offset shall be the constant $0$. Formally:
\begin{equation}
\mathsf{MME}.\mmbase(\cvar)\ \triangleq\ 
	\compile{(\logic{\mathabs{b}},0),\mathabs{\ell}} 
	\mbox{ where } 
    \mathsf{PA}.\mmvdfont{base}(\cvar)=\mathabs{\ell},
    \mmvdfont{domain}(\mathabs{\ell})=\{\mathabs{b}\}
\end{equation}

\noindent
The shifting of a location in $\mmloc$
	by an expression $e$ is computed based on the abstract shift operation 
as follows:
\begin{equation}
\mathsf{MME}.\mmshift(\mmv,\compile{(\logic{e_b},\logic{e_o}),\mathabs{\ell}}, e)\ \triangleq\ 
	\compile{(\logic{e'_b},\logic{e_o+e}),\mathabs{\ell}_s}
\end{equation}
where 	$\mathabs{\ell}_s=\mathsf{PA}.\mmvdfont{shift}(\mmv_s,\mathabs{\ell},e)$
and the new logic base $e'_b$ selects (using a conditional expression) the base $\mathabs{b}_i$ from the ones of $\mathabs{\ell}_s$. 
Let us denote by $\mathsf{fits}(e_b,\mathabs{\ell},\mathabs{b})$ the boolean term testing that the block identifier in $e_b$ is one of the blocks identifiers in $\mathsf{PA}.\mmvdfont{domain}(\mathabs{\ell})$ which has the same symbolic block (\ie, base) as $\mathabs{b}_i$, \ie:
\begin{equation}
\mathsf{fits}(e_b,\mathabs{\ell},\mathabs{b})\ \triangleq\ 
	\bigvee_{\mathabs{b}_j \in \mathsf{PA}.\mmvdfont{domain}(\mathabs{\ell})\textrm{ s.t. }\mathsf{PA}.\mmvdfont{base}(\mathabs{b}_j)=\mathsf{PA}.\mmvdfont{base}(\mathabs{b})} e_b = \mathabs{b}_j
\end{equation}
Using $\mathsf{fits}$, if 
$\mathsf{PA}.\mmvdfont{domain}(\mathabs{\ell}_s)$ is 
	$\{\mathabs{b}_1,\ldots,\mathabs{b}_n\}$,
the formal definition of $e'_b$ is:
\begin{equation}
e'_b \ \triangleq\ \left(\begin{array}{ll}
	\mathsf{fits}(e_b,\mathabs{\ell},\mathabs{b}_1) \land \mathsf{PA}.\mmvdfont{slice}(\mathabs{b}_1,\logic{e_o+e})
		\logic{~?~\mathabs{b}_1 ~:~} \\
	\quad{\ddots}\  
	\mathsf{fits}(e_b,\mathabs{\ell},\mathabs{b}_{n-1}) \land \mathsf{PA}.\mmvdfont{slice}(\mathabs{b}_{n-1},\logic{e_o+e})\logic{~?~\mathabs{b}_{n-1}~:~ 
	\mathabs{b}_n} 
	\end{array}\right)\label{eq:exp-shift}
\end{equation}
Indeed, since the shift operation can not change the symbolic block, we have to test, using $\mathsf{fits}$, that each resulting block identifier $\mathabs{b}_i$ has the same symbolic block as $e_b$.

The size of the expression encoding $\mathsf{MME}.\mmshift$ depends on the product of sizes of domains computed by $\mathsf{PA}$ for $\mathabs{\ell}$ and $\mathabs{\ell}_s$. 
If the abstract locations have a singleton domain, \ie\ 
$\mathsf{PA}.\mmvdfont{domain}(\mathabs{\ell}_s)=\{\mathabs{b}_1\}$, then $e'_b$ is simply $\logic{\mathabs{b}_1}$.
When the precision of the SA does not enable such simplification, 
we could soundly avoid big expressions generated by $\mathsf{MME}.\mmshift$ by using in $\mathsf{MME}.\mmload$ and $\mathsf{MME}.\mmstore$ operations only the component abstract location of an environment's location.

\mypar{Loading from memory:}
Reading an integer value in the environment $\mmv$ at a location 
$\mml=\compile{(e_b,e_o), \mathabs{\ell}}$ 
is compiled into a read operation (denoted by $a[e]$ for concision) from an array variable obtained by statically dispatching the logical base $e_b$ of $\mml$ among the possible base identifiers in $\mathsf{PA}.\mmvdfont{domain}(\mathabs{\ell})=\{\mathabs{b}_1,\ldots,\mathabs{b}_n\}$ as follows:
\begin{equation}
\mathsf{MME}.\mmload(\mmv,\carith,\compile{\logic{(e_b,e_o)}, \mathabs{\ell}}) 
	\triangleq\ \vint{e}
\end{equation}
where
\begin{equation}
e \ \triangleq\ \left(\begin{array}{ll}
	\logic{e_b=\mathabs{b}_1~?~}\mmv_\epsilon(\mathabs{b}_1)\logic{[e_o] ~:~} \\
	\quad{\ddots}\  
	\logic{e_b=\mathabs{b}_{n-1}~?~}\mmv_\epsilon(\mathabs{b}_{n-1})\logic{[e_o]  ~:~} 
	\mmv_\epsilon(\mathabs{b}_{n})\logic{[e_o]}
	\end{array}\right)\label{eq:exp-load}
\end{equation}
The size of the expression above may be reduced by asking to SA an over-approximation $\mathabs{o}$ of the values of expression $e_o$ in the current state. If SA is able to produce a precise result for $\mathabs{o}$, we could remove from the expression above the cases for abstract blocks $\mathabs{b}_j$ for which $\mathsf{PA}.\mmvdfont{slice}(\mathabs{b}_j,\mathabs{o})=\textit{false}$ (i.e., the formula is invalid for the values in $\mathabs{o}$).

The expression in equation (\ref{eq:exp-load}) is also used for
reading pointer values. In this case, the expression obtained is a
tuple. The abstract location corresponding to this logic expression is
obtained using the abstract $\mathsf{PA}.\mmvdfont{load}$ operation in
the abstract state component $\mmv_s$ of the environment:
\begin{equation}
\mathsf{MME}.\mmload(\mmv,\cptr{\cty},\compile{\logic{(b,o)}, \mathabs{\ell}}) \triangleq\
	\vptr{e,\mathsf{PA}.\mmvdfont{load}(\mmv_s,\cptr{\cty}, \mathabs{\ell})}
\end{equation}

\mypar{Storing in memory:}
The compilation of $\amstore$ semantic operation is done by the $\mathsf{MME}.\mmstore$ operation that produces a new environment $\mmv'$ and a boolean term (formula) $e'$ encoding the relation between the logic arrays associated to blocks before and after the assignment as follows:
\begin{equation}
\mathsf{MME}.\mmstore(\mmv,\cty,\compile{\logic{(e_b,e_o)},\absloc},\textsf{v}) 
	\triangleq\ \mmv',\logic{e}'\ 
	\mbox{ for }\ 
	\textsf{v} \in\{\mmvint(\logic{e}), \mmvptr(\compile{\logic{e},\absloc_v})\}
\end{equation}
where $\mmv'=\compile{\mathabs{s'},\mmv'_\epsilon}$ with 
$\mathabs{s'}$ the abstract state computed by the analysis for the control pointer after the assignment compiled.
The new block mapping $\mmv'_\epsilon$ uses fresh logic variables for the abstract blocks in the domain $\mathsf{PA}.\mmvdfont{domain}(\absloc)=\{\mathabs{b}_1,\ldots,\mathabs{b}_n\}$ of the abstract location $\absloc$ at which is done the update:
\begin{equation}
 \mmv'_\epsilon \triangleq\ \mmv[\mathabs{b}_1 \longleftarrow \alpha_1,\cdots, \mathabs{b}_n \longleftarrow \alpha_n]
\end{equation}
The fresh variables are related with the old ones using the store operator on logic arrays, denoted by $a[i\longleftarrow e]$, in the generated formula $e'$ defined as follows:
\begin{equation}
e'\ \triangleq\  
	\logic{\wedge_{i=1}^n \big(
		(e_b = \mathabs{b}_{i})~?~ 
			\alpha_i = }\;\mmv[\mathabs{b}_i]\logic{[e_o \longleftarrow e]
		~:~\alpha_i = }\;\mmv[\mathabs{b}_i]\logic{\big)}\label{eq:exp-store}
\end{equation}
The size of this expression may be reduced using the SA results in a similar way as for $\mmload$.
In general, the size of expressions generated by the compilation in 
	equations (\ref{eq:exp-shift}), (\ref{eq:exp-load}) and (\ref{eq:exp-store})
depends on size of the domain for the abstract locations computed by the static analysis. Indeed, if the analysis always provides abstract locations with a singleton domain, the compilation produces expressions with only one component, while proving most separation annotations.
However, if the analysis computes a small set $\mmvdblk$  
	(however bigger or equal to the number of program variables), 
the VC generated does not win any concision (we are falling back to the separation given by the typed model).

\mypar{Functor's properties:}
The requirements on the signature \textsf{PA} ensure that the operations $\mmvdfont{domain}$, $\mmvdfont{load}$ and $\mmvdfont{shift}$ are sound.
This enforces the soundness of definitions for the MME's operations.
Based on this observation, we conjecture\todohard{MS: proof idea?}{} 
that these operations compute a sound post-condition relation, although this relation maybe not the strongest post-condition. A formal proof is left for future work.

}




{

\abovedisplayskip=8pt plus 3pt minus 7pt
\belowdisplayskip=8pt plus 3pt minus 7pt

\section{Instances of Pointer Analysis Signature}
\label{sec:mm-with-ai}

The signature $\mathsf{PA}$ may be implemented by several existing pointer analyses. We consider three of them here and we show how they fulfill the requirements of $\mathsf{PA}$. We also define an analysis which exploits the results of a precise pointer analysis to provide an appropriate partitioning of the memory in $\mathsf{PA}.\mmvdblk$.

All pointer analyses we consider computes statically the possible values 
(i) of an address expression, \ie, an over-approximation of $\sem{\caexp}$ ($\caexp\in\caexps$ from Figure~\ref{fig:syntax}) and
(ii) of an address dereference, \ie, an over-approximation of $\sem{*\caexp}$.
For these reason, these analyses belong to the points-to analyses class~\cite{conf/paste/Hind01}.

\subsection{Basic Analyses ($\erbase$ and \erbasetop)}
\label{sec:points-to}

The first points-to analysis abstracts locations by a finite set of pairs 
$(\cvar,\mathabs{I})$ built from a symbolic block identifier $\cvar$ and 
an abstraction for sets of integers $\mathabs{I}$ collecting the possible offsets of the location in the symbolic block.
If we fix $\mathabs{\mathcal{I}}$ to be the abstract domain used to represents sets of integers, then the abstract domain for locations is defined by
$\absLoc \ \triangleq\ \pset{\cvars \times \mathabs{\mathcal{I}}}$.

Many abstract domains have ben proposed to deal with integer sets in abstract interpretation framework. For points-to analysis, most approaches use the classic domain of intervals \cite{popl/CousotC77}.
To obtain more precise results, we consider here the extension of the interval domain  which  also keeps  modulo constraints and small sets of integers.
This domain is implemented in the \Value\ plugin of \framac~\cite{fac/KirchnerKPSY15}. 
Then, the abstract sets in $\mathabs{\mathcal{I}}$ are defined by the following grammar:
\begin{equation}
\mathabs{\mathcal{I}}\ni\mathabs{I} ::= 
	\top \,\mid\, 
	[i_{\infty} .. i'_{\infty}] r\%n \,\mid\, 
	\{i_1, \ldots, i_n\} \label{eq:intset}
\end{equation}
where $r,n\in\NN$ are natural constants, 
$i_1,\ldots,i_n\in\ZZ$ are integer constants
and $i_{\infty},i'_{\infty}\in\ZZ\cup\{+\infty,-\infty\}$ are integer constants extended with two symbols to capture unspecified bounds.
We wrote $[i_{\infty} .. i'_{\infty}]$ for $[i_{\infty} .. i'_{\infty}]0\%1$.
The concretization of a value $\mathabs{I}$ in $\mathabs{\mathcal{I}}$, $\gamma:\mathabs{\mathcal{I}}\to\pset{\ZZ}$ maps 
$[i_{\infty} .. i'_{\infty}] r\%n$ to the set of integers $k\in[i,i']$ such that $k\%n=r$.
Because the abstract intervals are used to capture offsets in symbolic blocks which have a known size (given by the ABI), the concrete offsets are always bounded, but they may be very large.
We obtain independence of the ABI by introducing unspecified bounds for intervals and the $\top$ value.
For efficiency, the size of explicit sets 
$\{i_1, \ldots, i_n\}$ is kept bounded by a parameter of the analysis, denoted in the following $\erival$.
The domain $\mathabs{\mathcal{I}}$ comes with lattice operators (\eg, join $\sqcup\abstr$) and abstract transformers for operations on integers. Our work requires a sound abstract transformer for addition, $\mathabs{+}$.


\mypar{Precise offsets (\erbase):}
Let us consider a precise instance of such an analysis, i.e. field-sensitive and employing the abstract domain of intervals $\mathabs{\mathcal{I}}$ defined above. 
Let $\mathabs{S}$ be the abstract domain for program's states implemented in this analysis. This domain captures the abstract values for all program's variables. 
We denote by $\mathsem{\caexp}\abstr(\cstmt)$ the abstract location (in $\absLoc$) computed by the analysis for the address expression $\caexp$ at statement $\cstmt$. For address expressions typed as pointer to pointer types, the abstract value of the address expression $\mathsem{\csyn{*}\caexp}\abstr(\cstmt)$ is also an element of $\absLoc$ and computes the points-to information.\todohard{BY: load is partial}{}

The types and operations of $\mathsf{PA}$ are shown in Figure~\ref{fig:PA-B}.
The symbolic blocks are not partitioned, since $\mmvdblk \triangleq \cvars$.
Then, the slice for a block is the set of valid offsets for the symbolic block and the generated constraint is very simple.
Abstract locations are shifted precisely using the abstract transformer for addition in $\mathabs{\mathcal{I}}$. It is usually precise when $e$ is a constant. 
The soundness properties required by $\mathsf{PA}$ are trivially satisfied due to the simple form of abstract blocks' type and the soundness of operations on the abstract domains used.

\begin{figure}[t]
\vspace{-2eX}
\begin{center}
\begin{eqnarray}
\mathcal{L} \ \triangleq\ \absLoc 
	& & \mmvdst \triangleq \cstmts\to\mathabs{S} \label{eq:B-LS}
\\
	\mathcal{B}\ \triangleq\ \cvars && \label{eq:B-B}
\mmvdfont{base}(\cvar) 
	\triangleq \{ (\csyn{v}, \{0\})\} 
\quad
\mmvdsli(\cvar,e) 
 \triangleq  0 \le e < \csizeof(\typeof(\cvar)) \quad\quad
\\
\mmvdfont{domain}(\absloc) & \triangleq & \{\cvar \mid (\cvar,I\abstr)\in\absloc \} \label{eq:B-domain}
\\
\mmvdfont{shift}(\cstmt, \absloc, e) 
	& \triangleq & \sqcup_{(\cvar_k, I_k\abstr) \in \absloc}\abstr \{ (\cvar_k, I_k\abstr \mathrel{+\abstr} \sem{e}\abstr(\cstmt)) \}
\\
\mmvdload (\cstmt, \cptr \cty, \sem{\caexp}\abstr(\cstmt)) 
	& \triangleq & \sem{\csyn{*}\caexp}\abstr(\cstmt) \label{eq:B-load}
\end{eqnarray}
\end{center}
\vspace{-2eX}
\caption{Implementation of $\mathsf{PA}$ by analyses $\erbase$ and \erbasetop}
\label{fig:PA-B}
\vspace{-4eX}
\end{figure}

\mypar{Imprecise offsets (\erbasetop):}
We also consider an instance of the points-to analysis which is not field-sensitive.
For example, the \erbasetop{} analysis computes for $\sem{\code{&SORT.out2}}\abstr(\cstmt_{3})$,
where $\cstmt_{3}$ is the assignment at line 3 of listing in Figure~\ref{lst:main}(a), 
the set of abstract location $\{(\code{df}i,\top), \ldots, (\code{pf}j, \top) \mid 1 \le i \le 8, 1 \le j \le 4\}$.
The definition of the elements of the signature $\mathsf{PA}$ is exactly the one given in Figure~\ref{fig:PA-B}.

\subsection{Partitioning by Cells (\ercell)}

Analyzers that do not handle aggregate types (arrays and structs) decompose the symbolic blocks of variables having aggregate types 
into atomic blocks that all have a scalar type.
We call these blocks \emph{cells}.
For examples, the symbolic block of variable \code{pf} in Figure~\ref{lst:main}(b) is split into four cells
of type \code{pos_t}. 
%
For this analysis, the definitions for $\mathsf{PA}$ are those given in Figure~\ref{fig:PA-B} except for the type $\mmvdblk$ and the operations using this type $\mmvdsli$ and $\mmvddom$.
To define $\mmvdblk$, we first define the set $\Cells(\cty)$ of \emph{cells-paths} of type $\cty$ by induction on the syntax of $\cty$ as follows: 
\begin{eqnarray*}
\Cells (\cty) & \triangleq & \left\{
\begin{array}{ll}
\{\epsilon\} & \textrm{if } \cty \in \cstyps
\\
\bigcup_{1 \le i \le n} \cfld_i \cdot \Cells (\cty_i) & \textrm{if }
                                                         \cty \text{ is
                                                         the record type }\{\cfld_1 : \cty_1, \dots, \cfld_n : \cty_n\}
\\
\bigcup_{0 \le i < n} [i] \cdot \Cells (\cty_e) & \textrm{if } \cty
                                                    \text{ is the
                                                    array type } \cty_e[n]
\end{array}
\right.
\end{eqnarray*}
where the operator ``$\cdot$'' prefixes each path of its second operand by its first operand.
For a variable $\cvar$, we define $\Cells(\cvar) = \cvar \cdot \Cells(\typeof(v))$.
For example in Figure~\ref{lst:main})(b), 
$\Cells(\mathtt{df}) = \{\mathtt{df}\cdot[0], \dots,  \mathtt{df}\cdot[7]\}$.
Given a cell-path $c$, we denote by $\CellRange(c)$ the
range of offsets (in bytes) that correspond to the path and which is computed using ABI.
Then, we replace definitions in equation (\ref{eq:B-B}-\ref{eq:B-domain}) from Figure~\ref{fig:PA-B} by:
\begin{eqnarray*}
\mmvdblk & \triangleq & \{\Cells(\cvar) \mid \cvar \in \cvars \} 
\quad\quad 
\mmvdsli(\cvar\cdot c, e)\ \triangleq\ e \in \CellRange(c)  \\
\mmvddom(\absloc) & \triangleq &
  \{ \cvar\cdot c \in \mmvdblk \mid \exists i \in \NN, (\cvar, i) \in \gamma (\absloc)
  \land i \in r(c) \}  
\end{eqnarray*}
meaning that the slice of a cell-path is given by the range of bytes corresponding to the cell, and 
the domain of an abstract location is defined by enumerating all cells
that intersect with abstract location's abstract offsets.

\subsection{Partitioning by Dereference Analysis (\erpa)}
\label{ssec:slice}

We have seen in Section~\ref{ssec:functor} that the size of generated VC  strongly depends on two factors: the size of $\mmvdblk$ and the number of abstract blocks in the domain of abstract locations. 
This section defines an analysis which, based on the results of \erbase, aims to minimize these two factors while still producing sound results.
Roughly, the idea is to group cells that are accessed by a set of left
values which is upwards-closed w.r.t. the relation ``points-to'' computed by \erbase.
Therefore, two different abstract blocks will never be pointed-to by the same left value, \ie,
if the domains of abstract locations 
	$\sem{\csyn{*}a_1}\abstr(\cstmt_1)$ and 
	$\sem{\csyn{*}a_2}\abstr(\cstmt_2)$ 
	share an abstract block $\mathabs{b}$, 
then $\sem{a_1}\abstr(\cstmt_1)$ and $\sem{a_2}\abstr(\cstmt_2)$ belong to the same block.

For this, we define a partition $P$ of \emph{pointer-typed left-values} used by statements of the current context call using the equivalence relation $\simeq$ defined as follows.
We denote by $\absloc\downarrow_n$ the set of concrete locations
$\gamma(\absloc+\abstr 0) \cup \ldots \cup \gamma(\absloc+\abstr n-1)$. 
Then, two left-values appearing in some statements are related by $\simeq$ if the concretization of the abstract locations computed by $\erbase$ for their addresses on the corresponding statements overlap. Formally, for any left-values $\clval_1$ and $\clval_2$ used in statements $\cstmt_1$ \resp\ $\cstmt_2$, 
\begin{equation*}
\big(\mathsem{(\&\clval_1)^{\cty_1}}\abstr(\cstmt_1)\downarrow_{n_1}\big)
\ \bigcap\ 
\big(\mathsem{(\&\clval_2)^{\cty_2}}\abstr(\cstmt_2)\downarrow_{n_2}\big)
\neq\emptyset \implies (\clval_1, \cstmt_1) \simeq (\clval_2, \cstmt_2)
\end{equation*}
where $n_i=\csizeof(\cty_i)$.  
%
%
By definition, this relationship is reflexive and symmetric, and
we close it transitively. It is computed  by a simple iterative process on top of the results of $\erbase$ analysis.
For a given element $p\in P$, 
we compute the set of concrete locations pointing to left-values in $p$:
%
\begin{equation*}
B(p)\ \triangleq\ \bigcup_{(\clval_i,\cstmt_i)\in p} \gamma(\mathsem{\&\clval_i}\abstr(\cstmt_i))
\end{equation*}
%
%
Analysis \erpa\ implements signature $\mathsf{PA}$ using the definitions in Figure~\ref{fig:PA-B} except for (\ref{eq:B-B}-\ref{eq:B-domain}) that are replaced by:
\begin{eqnarray*}
\mmvdblk & \triangleq & \{\compile{\cvar, s} \mid \exists p \in P \land s=\{i \mid (\cvar, i) \in B(p)\} \} 
\\
\mmvdsli(\compile{\cvar, s}, e) & \triangleq & e \in s\quad\quad \\
\mmvddom(\absloc) & \triangleq &
  \{ \compile{\cvar, s} \in \mmvdblk \mid \exists i \in \NN, (\cvar, i) \in \gamma (\absloc)
  \land i \in s \}
\end{eqnarray*}
In the example on Figure~\ref{lst:main}(b), if $\erbase$ is precise enough,
$\erpa$ computes a $\mmvdblk$ which splits the symbolic block labeled by the array variable \code{df} into (only) two abstract blocks: 
one for the bytes located at indexes $[1 .. 4]$ (whose addresses are stored in input fields)
and another for indexes $\{0\}\cup[5 .. 7]$ (stored in output fields).
%
%

}




\section{Experimental Results}
\label{sec:experiments}




\subsection{Implementation}


We implemented our framework in \framac \cite{fac/KirchnerKPSY15}, an extensible and modular platform for the analysis of software written in C.
\framac\ includes various plug-ins, interacting with each other through interfaces defined by the platform. 

The plug-in \Value{} is a context-sensitive static analyzer based on abstract interpretation; it employs several numerical abstract domains, including the one defined in eq.~(\ref{eq:intset}) for sets of integers. 
On top of the value analysis provided by \Value{}, which includes the \erbase\ analysis from Section~\ref{sec:mm-with-ai},
we coded new partition analyses to obtain analyses \erbasetop, \ercell\ and  \erpa.

The \WP{} plug-in of \framac\ is a DV tool which also includes  
a built-in simplifier for formulae, \Qed~\cite{nfm/Correnson14}, a driver to call SMT solvers and 
the signature \textsf{MME} for memory model environments~\cite{WPreport}.
We coded in \WP{} the signature \textsf{PA}, the functor defined in Section~\ref{ssec:functor}, and each implementation of \textsf{PA} for the above static analyses.
The full development represents 1680 LoC of \textsf{Ocaml}.

\subsection{Experimental setup}

\noindent
\emph{Case study:}
We consider a case study which extends our running example from Figure~\ref{lst:sort4} such that the type \code{data_t} is a record which encapsulates numerical values to be sorted and other information. 
We attempt to prove the functional correctness of the \code{sort} function for various number of inputs $\ersigs \in \{4, 8, 16, 32\}$.
The specification of \code{sort} consists of 40 ACSL properties, which \WP{} transforms into 62 VC 
for each memory model.
%
We also consider 3 different context calls for \code{sort} as the entry point for the analysis. They initialize the fields of the \code{SORT} variable using pointers to:
variables on the stack similar to Figure~\ref{lst:main}(a) (\ervars),
fields of a single record (\erstruct) and 
two arrays (for values and permutations) (\erarrays).
In addition, we consider two variants for contexts \erstruct{} and \erarrays{}. 
In the (\erreg) variant, all input and output fields are grouped together, \ie, inputs point to the first $\ersigs$ fields/indexes in a regular way and outputs to the remainder.
For the (\errdm) variant, inputs and outputs are initialized in 
a randomized order, as in Figure~\ref{lst:main}(b) for \erarrays.
The latter case is designed to defeat points-to analyses where offsets
are abstracted solely by intervals plus congruences.

\mypar{Variants of memory models:}
For comparison with the basic DV tools, we also conduct proof using the default memory model of \WP{} (case $\ertyped$). 
To observe the influence of the precision of points-to analysis \erbase\ on the generated memory models environments, we vary the parameter $\erival$ which gives the upper limit for the size of small sets kept by the abstract domain $\mathabs{\mathcal{I}}$ in Section~\ref{sec:points-to}.
We apply \erbase\ for $\erival$ in $\{4, 8, 16, 32\}$ to generate its  memory model environment and the VC. 
For the same values, we launch the \ercell\ (\resp\ \erpa) analysis after  \erbase\ and generate the corresponding environments.

\mypar{Proving VCs:}
\WP{} generates VC using the library for many sorted first-order logic provided by \Qed. After applying on-the-fly\todoeasy{important? classic term? =syntactic}{} simplifications of VCs, \Qed\ exports the
VC to back-end solvers. We configure \WP{} to discharge simplified VCs to the Alt-ergo prover and the remaining unproved VCs to be sent to CVC4.
Those experiments ran on an Intel(R) Xeon(R) CPU E5-2660 v3 @ 2.60GHz with a
timeout of 10 seconds per goal for each solver.

\vspace{-1eX}
\subsection{Results}
\begin{figure}[t]
\vspace{-2eX}
    \centering
    \includegraphics[width=1\textwidth,trim=10mm 10mm 10mm 7mm]{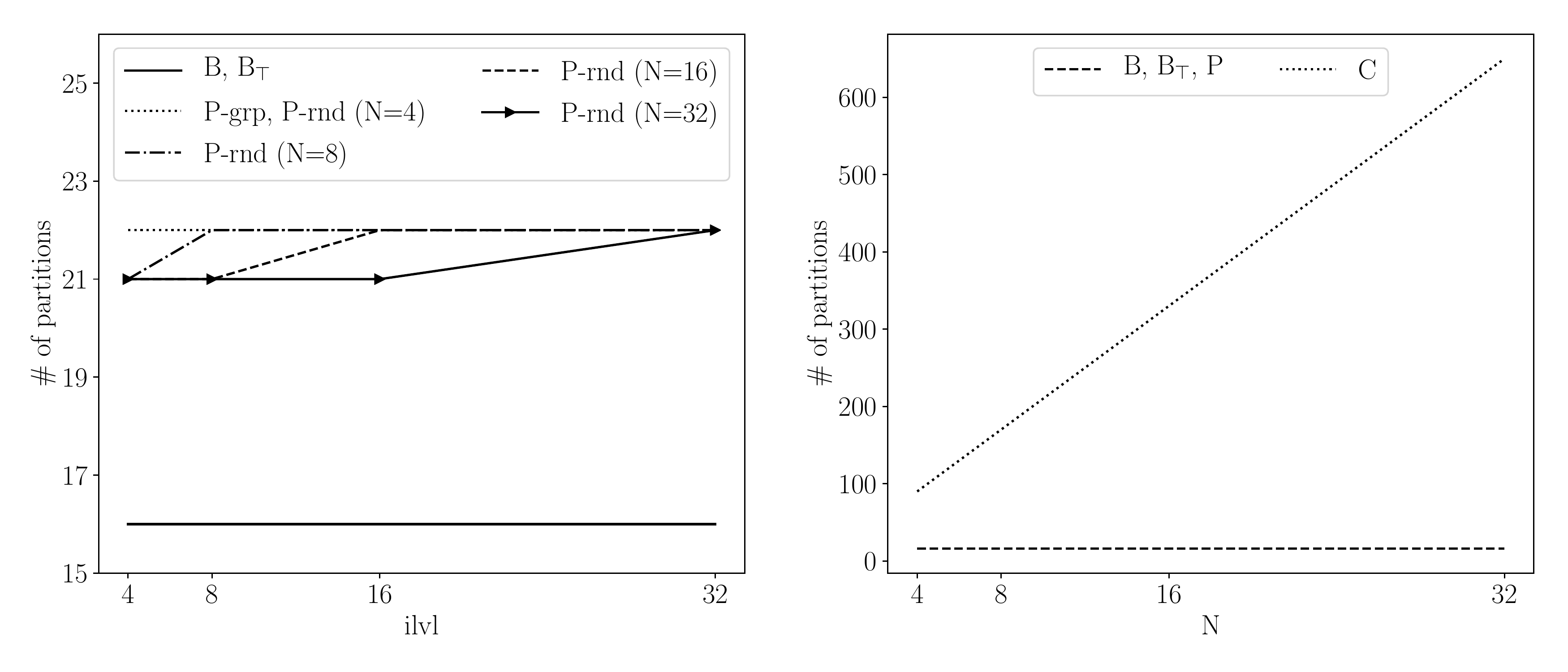}%
\vspace{-2eX}
\caption{Comparison between analyses on number of partitions}
\label{fig:part-ival}
\vspace{-4eX}
\end{figure}

Figure~\ref{fig:part-ival} shows the number of partitions (size of $\mathsf{PA}.\mathcal{B}$ ) inferred by the various analysis for a given call context.
Recall that the partitioning generated by $\erbase$ is always constant, since fixed by the program variables.
As expected, $\ercell$'s result is linear in the number of inputs (right plot in ~\ref{fig:part-ival}).
The partitioning by $\erpa$ creates fewer abstract blocks
when $\ersigs$ is less than $\erival$ (left plot in ~\ref{fig:part-ival}).
\todoeasy{Not clear why}{}
Fewer blocks means a less precise analysis: in our example,
the two equivalence classes that get merged are those corresponding
to inputs and outputs.\todoeasy{Not clear}{}

\begin{figure}[t]
\vspace{-2eX}
\begin{center}
\includegraphics[width=0.5\textwidth,trim=10mm 5mm 0mm 5mm]{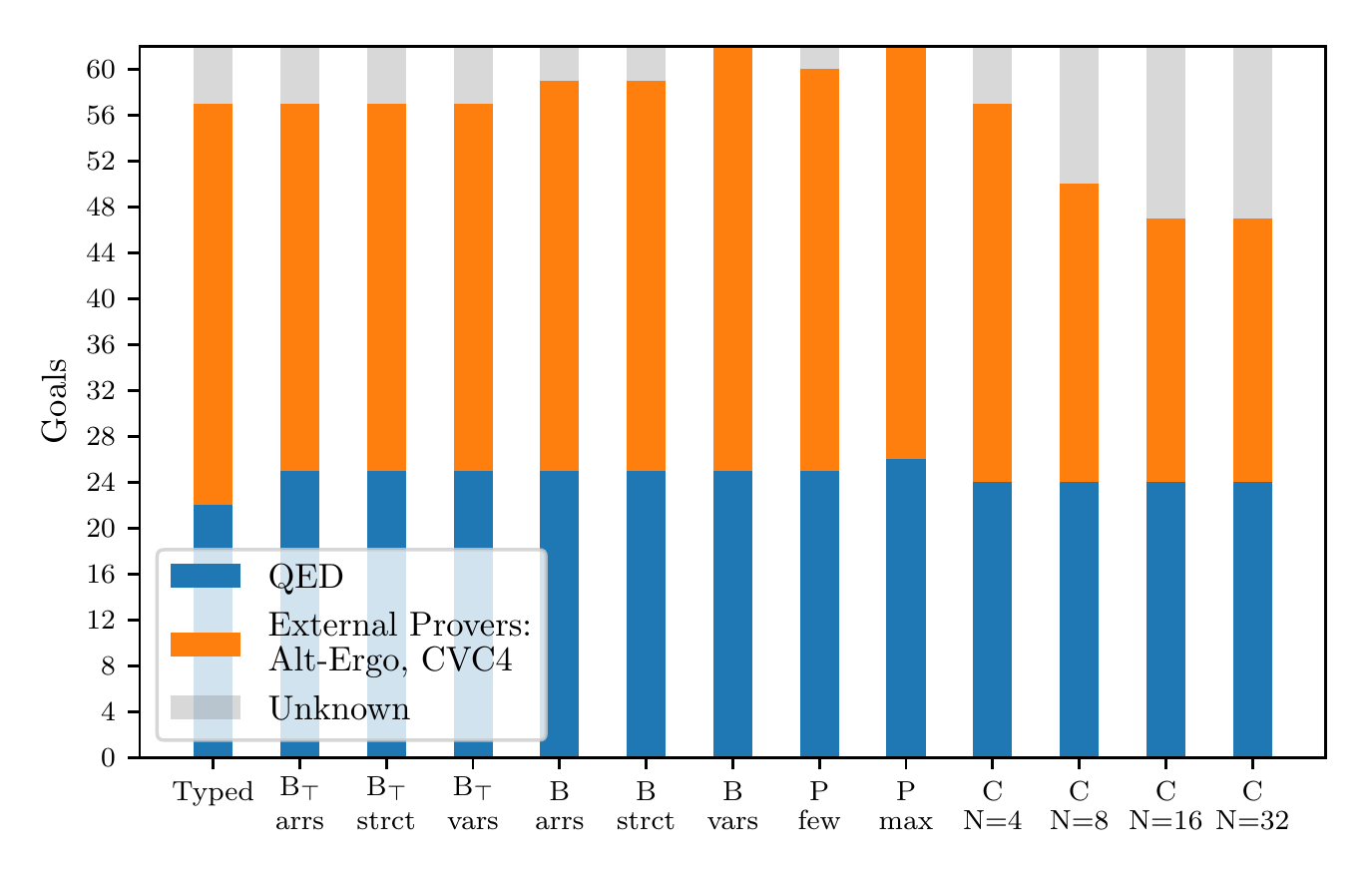}%
\includegraphics[width=0.5\textwidth,trim=7mm 10mm 10mm 10mm]{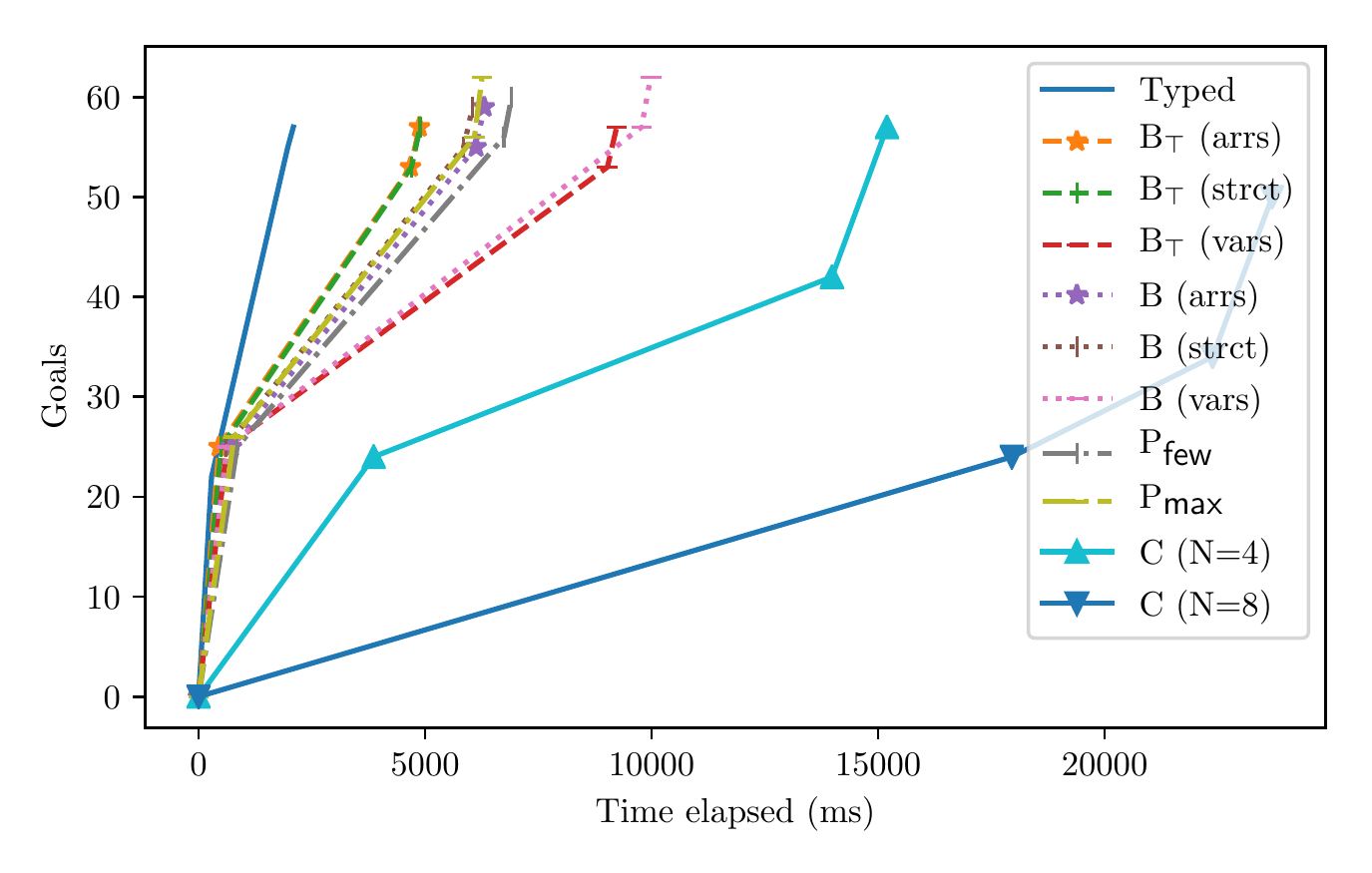}%
\end{center}
\vspace{-2eX}
\caption{Results for solving generated VC}
\label{fig:exps}
\vspace{-4eX}
\end{figure}

Figure~\ref{fig:exps} (left) shows that $\erbase$~partitioning is sufficient to prove all goals for the $\ervars$ context, since all
values are implicitly separated onto different symbolic bases.  However, for
contexts $\erstruct$ and $\erarrays$, inputs and outputs share the same symbolic base which is too imprecise to prove all goals. 
Analysis \erbasetop\ infers that the fields of \code{SORT} point
to all possible inputs and outputs, which yields even worse results.
The results for $\ercell$~partitioning worsen with the increase in the number of inputs due to the complexity of the VCs generated. 
For $\erpa$~partitioning, we are not interested in the $\ervars$ context
considering it is but a small refinement of $\erbase$ in that context. In our
experiments, we were able to identify two classes of experiments giving
similar results in term of
provability and time: $\erpamax$ are results for experiments where partitions
are maximal and conversely for $\erpafew$.
For readability reasons, we display only the worse results of those two classes.

Figure~\ref{fig:exps} (right) shows the total time spent on the VCs that get proven
(\ie, do not timeout),
for each model. We observe that more partitions lead to bigger VCs which
takes more time to be proven, especially for $\ercell$ partitioning.
Refining $\erbase$ partitioning within $\erpa$ leads to a better provability
at the cost of a negligible increase in time in provers.
These results demonstrate that $\erpa$ analysis offers the best trade-off
between partition's granularity and provability in reasonable time,
regardless of the context. 
Moreover, all verification conditions are proved for the regular context;
for randomized contexts, better results are obtained by increasing the precision of points-to analysis \erbase.
The improvement is real because $\erbase$ exhibits the same performance only for the $\ervars$ context.



\vspace{-2eX}
\section{Related Work and Conclusion}
\label{sec:rw}

\vspace{-1eX}
\noindent
\emph{Memory Model for C:}
Program verification and certified compilation have proposed several memory models to capture the semantics of C pointers.
All these models view the memory as a collection of disjoint regions.
%
Two main classes may be distinguished: 
(i) the regions are typed by the value stored, therefore regions storing values of different types are disjoint
and
(ii) the regions are seen as raw arrays of bytes to capture low-level manipulations of memory in C.
The first class provides a good abstraction for verification of type-safe languages,
 (e.g., Java-like~\cite{barnett2004spec,barnett2005boogie}, HOL~\cite{DBLP:conf/cade/MehtaN03}) 
 	or type-safe C programs
	(GRASSHoper~\cite{Piskac2014}, HIP/Sleek~\cite{CHIN20121006}).
The second class is mainly used inside 
	static analyzers for C 
	(Infer~\cite{CaDO2006}, 
	MemCAD~\cite{ChangRN07}, 
	\Value{}~\cite{Buhler17}) or 
	deductive verifiers 
	(Caduceus~\cite{DBLP:conf/cav/FilliatreM07},
	HAVOC~\cite{journals/sttt/ChatterjeeLQR09},
 	SMACK~\cite{RakamaricH09},
	VCC~\cite{DBLP:conf/cade/BohmeM11}, VeriFast~\cite{Jacobs}).
Hybrid memory models  
	either introduce typing in raw memory models for efficiency,
	or introduce raw models in typed ones for precision.
\WP{} supports both classes of models and provides instances of the environment $\textsf{MME}$ for them \cite{WPreport}.

The CompCert project~\cite{jar/LeroyB08,inria/LeroyAB12} employs an abstract memory model to capture in an uniform way refinements of memory models for the certified compilation of C.
This work also inspired \cite{DBLP:journals/entcs/SotinJR10}, which surveys several concrete memory models for C and proposes a method to design static analyzers based on abstract memory models.
\Value{} is not built following these principles for efficiency reasons.

\mypar{Separation Logic versus FOL:}
Separation Logic~\cite{OHearnRY01} is used in many verification tools for C (e.g., GRASSHoper, HIP/Sleek, Infer, VCC, VeriFast) due to the efficiency of local reasoning. The specification logic used in \framac, ACSL~\cite{ACSL}, includes a separating conjunction operator (understood by \WP{} and \Value{} plugins), but it is far weaker than the standard separating conjunction operator.
The underlying solvers for SL of the above tools are either not available or deal with the type safe fragment of C. The recent SL-COMP initiative motivated the development of several independent solvers for type safe fragments of SL, one of them included in the CVC4~\cite{ReynoldsIS16} solver.
Our work focuses on DV tools using FOL and infers separation properties between memory regions. Our pointer analyses may be used in SL-based tools to obtain precise properties on arrays of pointers.\todowarn{really?}{}

\mypar{Pointer Analyses for DV:} 
Static analysis based on region inference is used in~\cite{hubert07hav} to partition a typed 
memory model.
The analysis is less precise than the points-to analysis in \Value{}
because the loss of precision for one location could force many precise locations to be collapsed in the same region.
\cite{RakamaricH09} employs pointer analysis to ensure a sound usage of the typed memory model in presence of casts. This work may be applied to extend the class of programs we deal with, but our focus is on improving efficiency of DV, not its realm.
Recent work \cite{DBLP:conf/vmcai/0062BW17} proposes a precise
points-to analysis to infer separation information in order to decrease the size of VCs. Although \Value{} is doing a less precise analysis, it is still able to infer such separation properties. In addition, we define a formalized channel to transfer such information to DV tools.
The authors of \cite{DBLP:conf/cade/BohmeM11} explore different memory models to generate with VCC a benchmark of problems for SMT solvers.
By implementing various memory models for \WP, we increase such benchmark.

\mypar{Conclusion:}
We have formalized the collaboration of a pointer analysis tool and a deductive verification tool by a functor which exploits the results of the pointer analysis to define sound and precise memory model environments used in the generation of verification conditions in first order logic theories. 
We applied this functor to several pointer analyses, including classic analyses (points-to analysis) and a new analysis that allows to obtain precise partitioning information of the program's memory.
We reported on the implementation of the functor in \framac\ and on the results obtained by different analyses on a benchmark of~C programs that exhibit complex features of pointers in C (arrays of pointers, duality of fields) and complex separation annotations. The results obtained show the interest of our functor for the automatization of deductive verification. 


\newpage
\bibliographystyle{abbrv}
\bibliography{biblio}

\appendix

\newpage
\section{Concrete Semantics and its Abstract Signature}
\label{sec:sem}

We provide here the standard semantic domain (concrete memory model) for the fragment of \clight~\cite{jar/BlazyL09} in the grammar on Figure~\ref{fig:syntax}. Then, we establish the correspondence between this concrete domain model and the abstract signature presented in Section~\ref{ssec:amm}.

The standard semantic domain of \clight\ is store-based, see Figure~\ref{fig:ssem}.
A stack ($\epsilon\in\textit{Stack}$) is a mapping from the program variables ($\cvar$) to the memory blocks ($\amblk$) where their content is stored. 
A store ($\sigma\in\textit{Store}$) is a mapping from addresses 
$\textit{Addr}=\amblk\times\textit{Offset}$ to values $\textit{Values} = \textit{Scalar}\cup\amloc$.
In our work, the scalar values (\textit{Scalar}) are integer values. 

\begin{figure}[b]
$\begin{array}{rlll}
s\in & \textit{State} & = \textit{Stack}\times\textit{Store}
	& \mbox{: program states}
\\
\epsilon\in & \textit{Stack} & = \cvars \ft \amblk
	& \mbox{: program stack}
\\
\sigma\in & \textit{Store} & = \textit{Addr} \ft \ZZ \cup \amloc
	& \mbox{: store mapping addresses to values}
\\
& \textit{Addr} & = \amblk \times \textit{Offset}
	& \mbox{: addresses}
\\
b\in & \amblk & & \mbox{: memory block identifier}
\\
\ell\in & \amloc & = \amblk \times \textit{Offset} 
	& \mbox{: standard interpretation of locations}
\\
o\in & \textit{Offset} & = \ZZ
	& \mbox{: numerical offsets}
\end{array}$
\caption{Concrete memory model}
\label{fig:ssem}
\end{figure}

Concrete memory models for \clight~\cite{DBLP:journals/entcs/SotinJR10} differ on the nature of pointers/locations ($\amloc$) and the nature of offsets (\textit{Offset}). 
Note that memory blocks have an unique identifier, but not a numerical address. Moreover, they are unordered. Therefore, it is impossible with an address $(b,o)$ to obtain the element of an other block $b'\ne b$, whatever the value of $o$. 

In \clight\ and other C-like low-level languages, everything that can be addressed can also be stored in a pointer. Therefore, $\amloc=\textit{Addr}$.
For the offsets, the lower-level memory model deals with true offsets (in bytes) within blocks, thus $\textit{Offset}=\ZZ$. This memory model can be used only when the target architecture is known (size of types, layouts). An Application Binary Interface (ABI) should provide such information, allowing an architecture-based manipulation of the numerical memory model.

The abstract signature in Figure~\ref{fig:am} provides:
\begin{itemize}
\item stack $\epsilon$ by the operator $\ambase$ which returns a location $(b,0)$ where $b=\epsilon(v)$ for each program variable $v$,
\item store $\sigma$ by the type $\ammem$ and the operations $\amload$ and $\amstore$,
\item in absence of dynamic allocation, block identifiers are given by program variables $\cvars$, 
\item locations by the type $\amloc$,
\item operations on locations by $\amshft$.
\end{itemize}


\newpage
\section{Detailed Experimental Results}

\begin{figure}
\vspace{-2eX}
    \centering
\includegraphics[width=0.47\textwidth,trim=10mm 0mm 0mm 0mm]{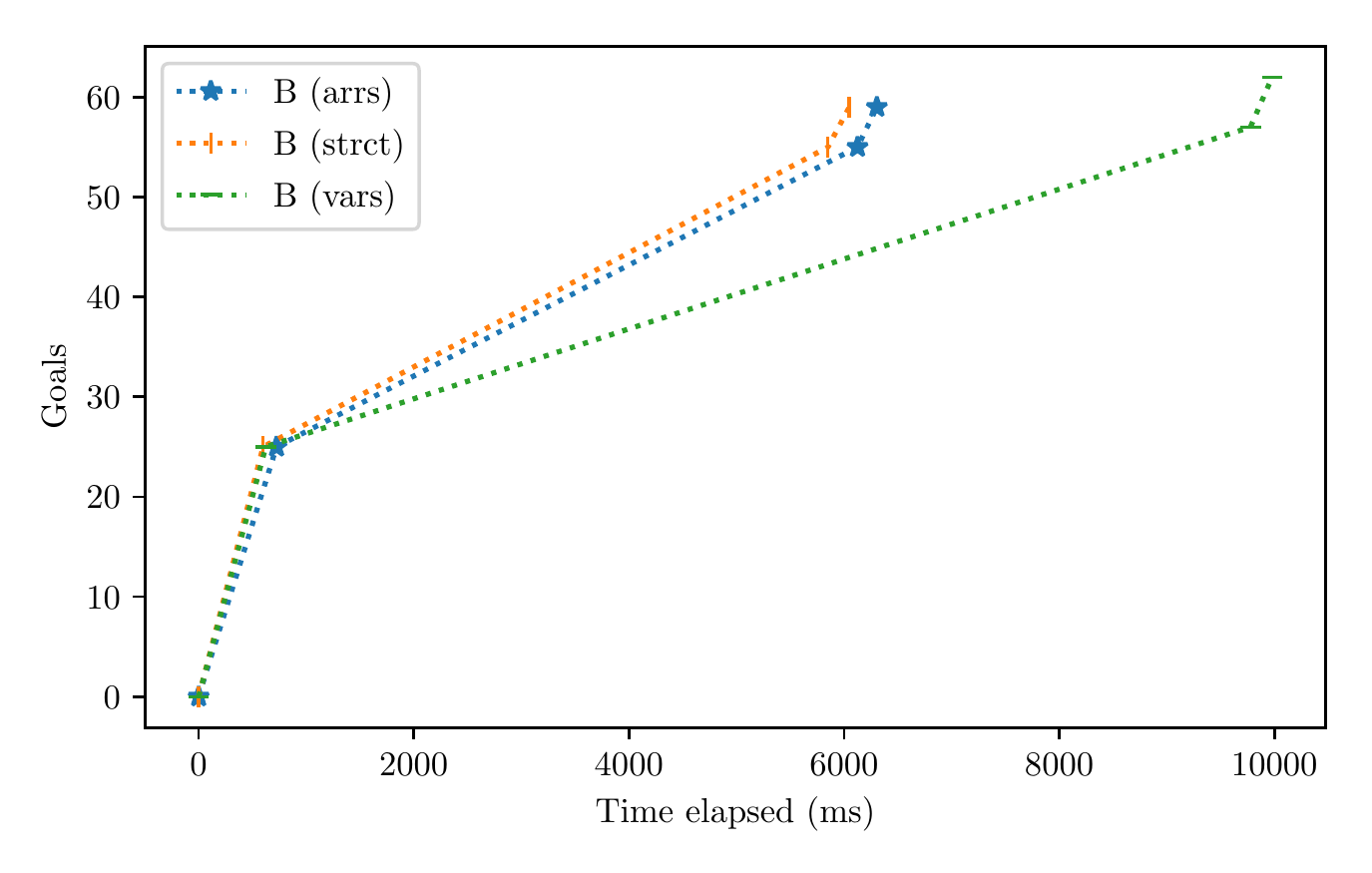}
\includegraphics[width=0.47\textwidth,trim=10mm 0mm 0mm 5mm]{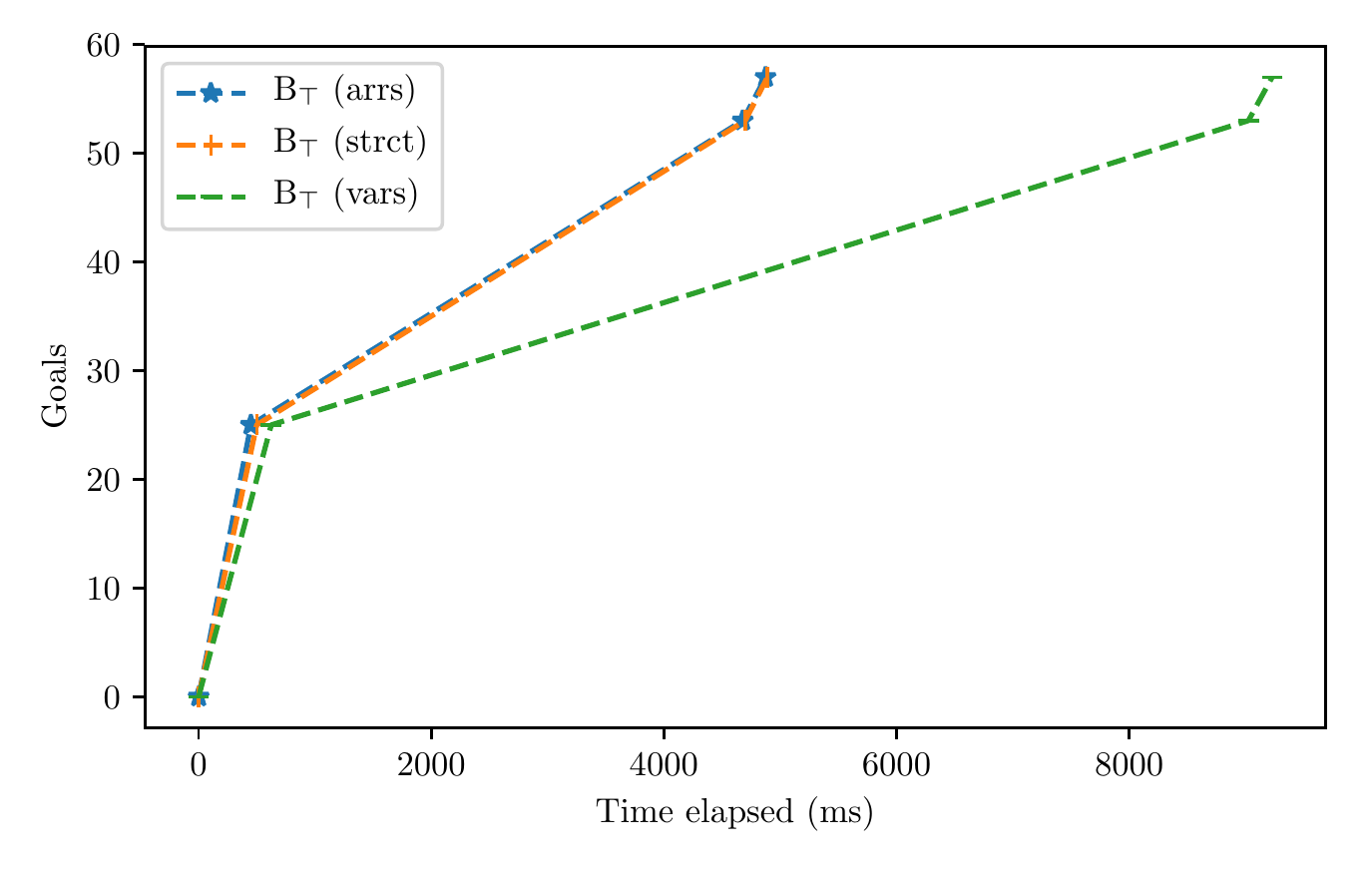}
\caption{Execution time and VC proved for memory models produced by \erbase\ \resp\ \erbasetop\ with different initialization contexts}
\vspace{-4eX}
\end{figure}

\begin{figure}
\vspace{-2eX}
    \centering
	\includegraphics[width=1\textwidth]{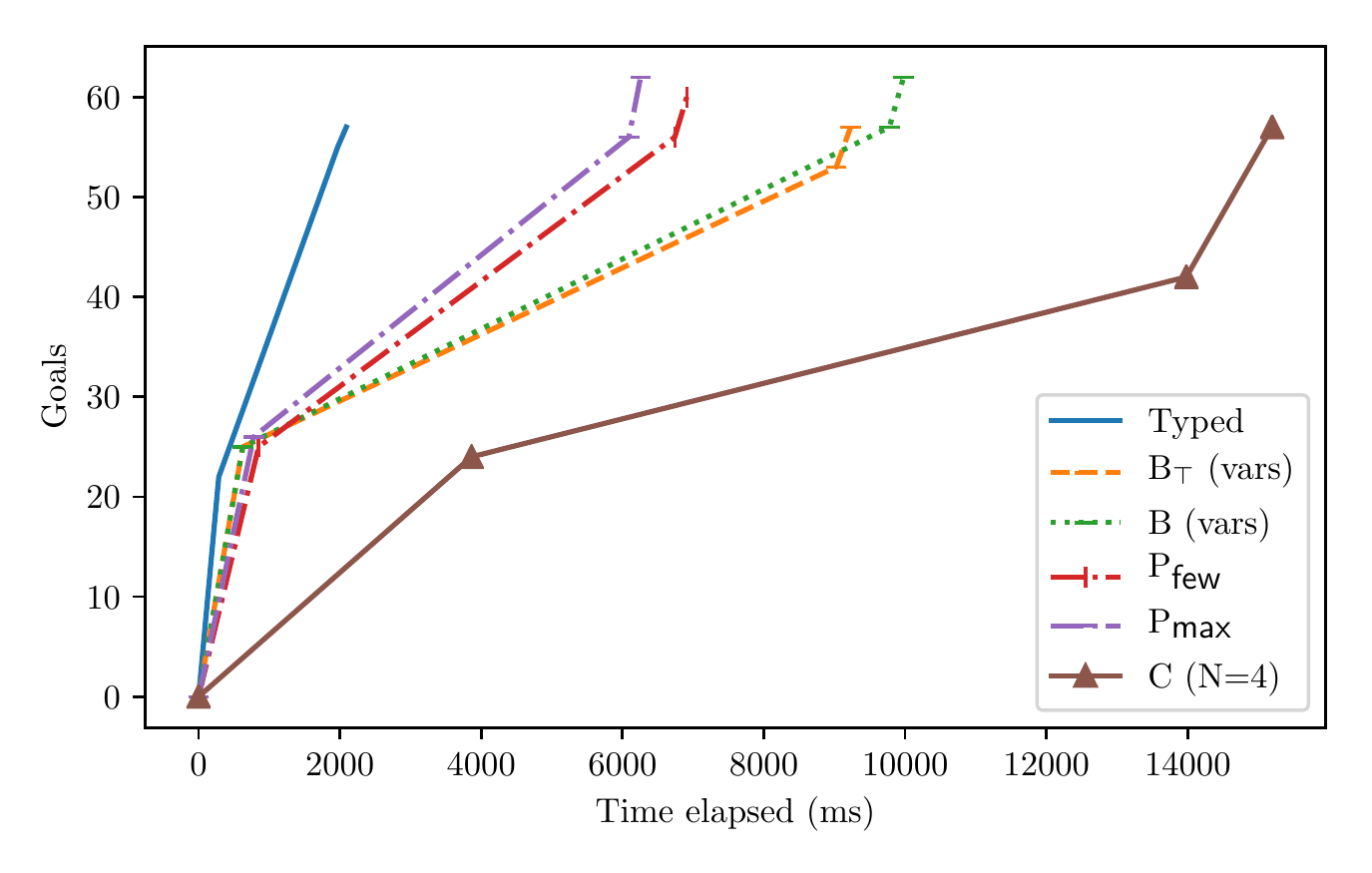}
\vspace{-4eX}
	\caption{Execution time and VC proved for initialization context using variables}
\vspace{-4eX}
\end{figure}

\begin{figure}
\vspace{-2eX}
    \centering
	\includegraphics[width=1\textwidth]{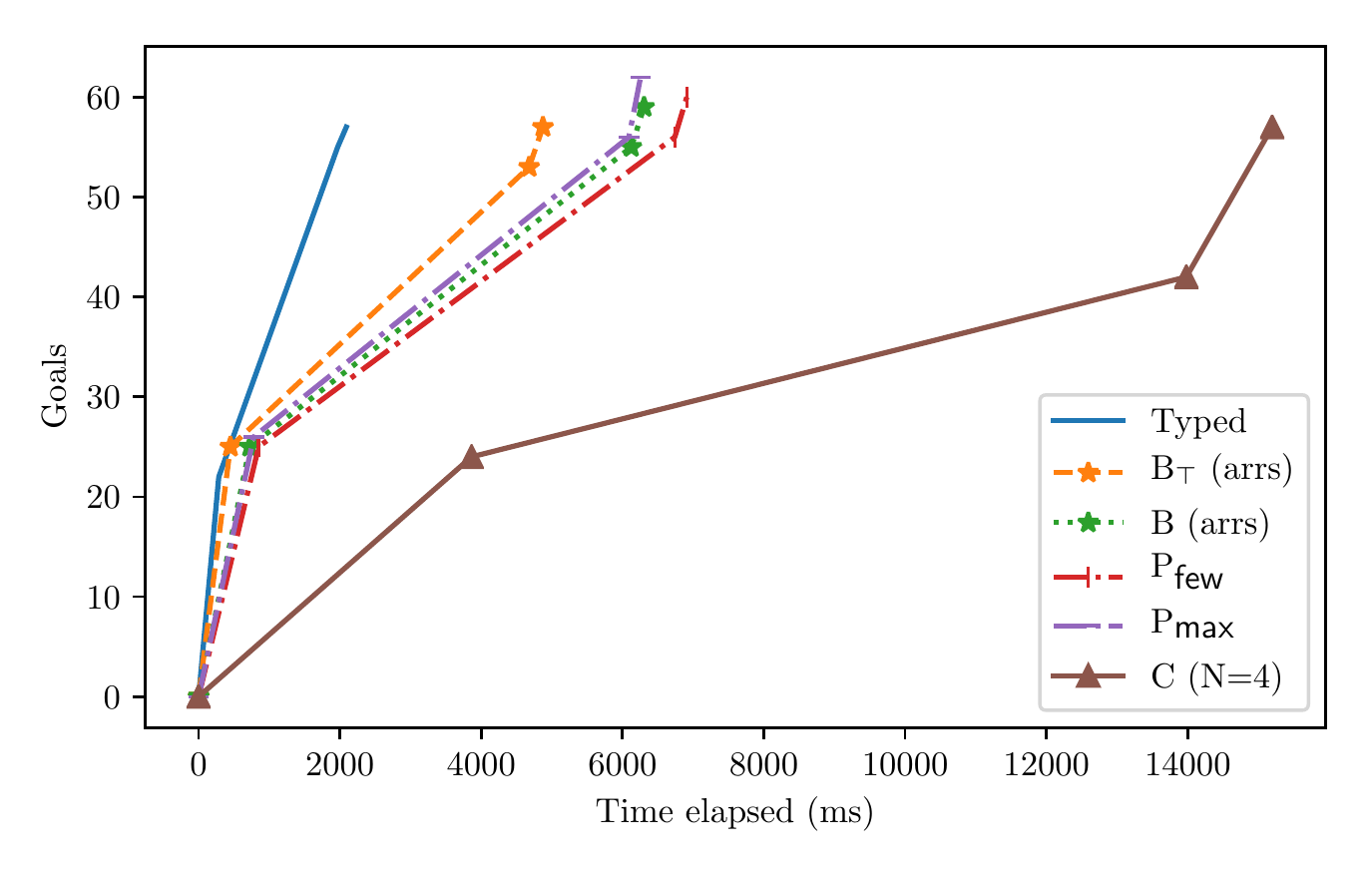}
	\vspace{-4eX}
	\caption{Execution time and VC proved for initialization context using arrays}
\vspace{-4eX}
\end{figure}

\begin{figure}
\vspace{-2eX}
    \centering
	\includegraphics[width=1\textwidth]{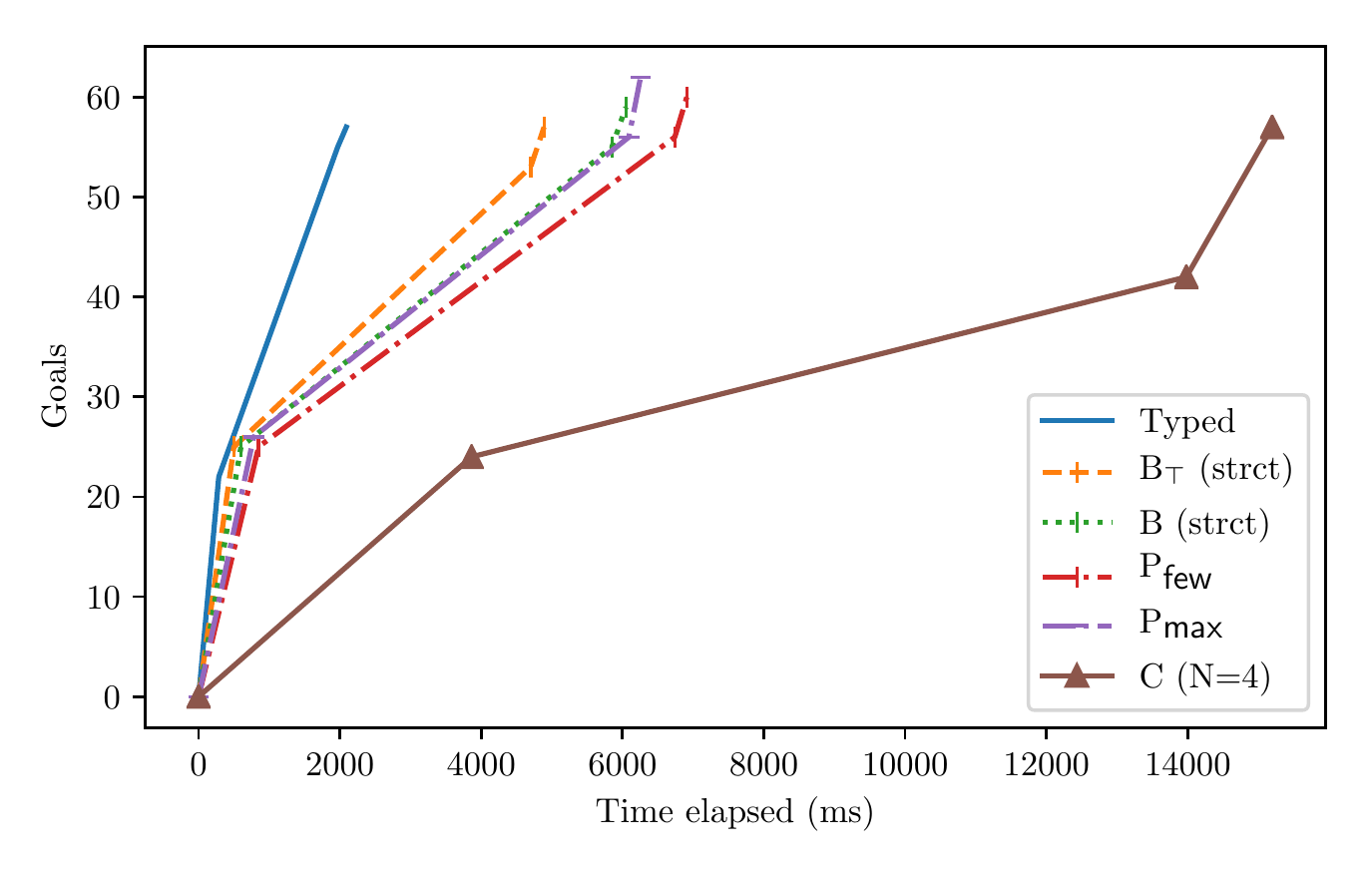}
	\vspace{-2eX}
	\caption{Execution time and VC proved for initialization context using records}
\vspace{-4eX}
\end{figure}

\end{document}